%% file: gausslectronitegrate.tex
\documentclass[aps,prb,twocolumn,10pt,superscriptaddress]{revtex4-2}

\usepackage{amsfonts}
\usepackage{amsmath}
\usepackage{amssymb}
\usepackage{mathtools}
\usepackage{tikz}
\usepackage{braket}
\usepackage{algorithm}
\usepackage[noend]{algpseudocode}
\usepackage{varwidth}
\usepackage{graphicx}
\usepackage[caption=false]{subfig}
\usepackage{dsfont}
\usepackage{accents}
\usepackage{xcolor}
\usepackage{hyperref}
\usepackage{orcidlink}

\DeclarePairedDelimiter\abs{\lvert}{\rvert}
\DeclarePairedDelimiter\norm{\lVert}{\rVert}
\DeclarePairedDelimiter\coul{(}{)}

\DeclareMathOperator*{\argmin}{arg\,min}
\DeclareMathOperator{\e}{e}
\DeclareMathOperator{\erf}{erf}

\algnewcommand{\LineComment}[1]{\State \(\triangleright\) #1}

\newcommand{\gridindex}[1]{\mathfrak{i}_{\Lambda_{#1}}}
\newcommand{\eriindex}[1]{\mathfrak{e}_{\Lambda_{#1}}}

\newcommand{\ud}{\mathrm{d}}

\newcommand{\N}{\mathbb{N}}
\newcommand{\Z}{\mathbb{Z}}
\newcommand{\R}{\mathbb{R}}
\newcommand{\C}{\mathbb{C}}

\newcommand{\addgets}{\stackrel{+}{\gets}}

\newcommand{\boxtrans}{\text{boxtrans}}
\newcommand{\orb}{\text{orb}}
\newcommand{\Oh}{\text{O}_{\text{h}}}
\newcommand{\Dfour}{\text{D}_4}

\graphicspath{{figures/}}
\makeatletter
\def\input@path{{figures/}}
\makeatother

\date{October 2026}

\begin{document}

\title{Computing electron overlap integrals for Gausslet orbitals on cubic lattices}

\author{Xianrui Yin~\orcidlink{0009-0008-6062-2625}}
\email{xianrui.yin@tum.de}
\affiliation{Technical University of Munich, CIT, Department of Computer Science, Boltzmannstra{\ss}e 3, 85748 Garching, Germany}

\author{Fereshteh Ghasempour~\orcidlink{0009-0007-5334-4068}}
\email{go56bew@mytum.de}
\affiliation{Technical University of Munich, CIT, Department of Computer Science, Boltzmannstra{\ss}e 3, 85748 Garching, Germany}

\author{Christian B.~Mendl~\orcidlink{0000-0002-6386-0230}}
\email{christian.mendl@tum.de}
\affiliation{Technical University of Munich, CIT, Department of Computer Science, Boltzmannstra{\ss}e 3, 85748 Garching, Germany}
\affiliation{Technical University of Munich, Institute for Advanced Study, Lichtenbergstra{\ss}e 2a, 85748 Garching, Germany}

\begin{abstract}
Gausslet orbitals on a cubic lattice, introduced in [\href{https://doi.org/10.1063/1.5007066}{Steven R.~White, J.~Chem.~Phys.~147, 244102 (2017)}], represent a localized, smooth, and systematically refineable basis set featuring a ``diagonal'' approximability of the electron repulsion integral tensor. The present work develops efficient algorithms for evaluating overlap integrals required for electronic-structure simulations using these Gausslets, specifically the kinetic, nuclear, and electron repulsion integrals (both with and without the diagonal approximation). Computational efficiency improvements rest on the exploitation of translation, permutation, and octahedral symmetries, a reordering and precomputation of nested sums, as well as an early truncation of small coefficients. Our algorithms reduce the number of electron repulsion integrals on a $5 \times 5 \times 5$ grid from $125^4 = 244140625$ to $138187$ due to symmetries, and achieve a wall-clock runtime for evaluating the remaining integrals with a truncation tolerance of $10^{-5}$ in under 1 second on a laptop computer. We apply the developed methodology to compute the ground state of the hydrogen atom and molecule as a demonstration.
\end{abstract}

\maketitle

\section{Introduction}

The publication \cite{White2017} by Steven White introduced a class of localized orbital basis sets defined on Cartesian lattices, denoted \emph{Gausslets}. A motivation for this work is their suitability for tensor network simulations, owing to desirable properties such as localization, orthogonality, smoothness, and diagonal approximability of the electron-repulsion integral tensor. One particular feature, which we will elucidate below, is the systematic refinability of the basis by a uniform rescaling.

As for any orbital basis set, their employment in electronic structure simulations necessitates the (pre-) computation of overlap integrals originating from the kinetic and Coulomb terms of molecular Hamiltonians. Reaching chemical accuracy with Gausslet orbitals on a uniform cubic lattice requires high resolution and, correspondingly, a large number of grid points, rendering overlap-integral computations challenging. In this work, we develop numerical algorithms for this purpose. Key ingredients for efficiency gains are exploiting translation, permutation, and octahedral symmetries, as well as reordering and precomputing nested sums.

As a toy-model illustration for such a precomputation, consider the expression
\begin{equation}
E \coloneq \sum_{\alpha=0}^{m-1} \sum_{\beta=0}^{m-1} \sum_{\gamma=0}^{m-1} c_{\alpha} \, c_{\beta} \, c_{\gamma} \, f(\alpha + \beta + \gamma),
\end{equation}
with prescribed coefficients $\{c_{\alpha}\}$ and some function $f$. The costly ``naive'' evaluation of the nested sums with complexity $\mathcal{O}(m^3)$ can be avoided by precomputing
\begin{equation}
t_{\kappa} \coloneq \sum_{\alpha=0}^{m-1} \sum_{\beta=0}^{m-1} c_{\alpha} \, c_{\beta} \, \delta_{\alpha + \beta, \kappa}, \quad \kappa = 0, \dots, 2(m - 1),
\end{equation}
and then evaluating, with asymptotic complexity $\mathcal{O}(m^2)$,
\begin{equation}
E = \sum_{\kappa=0}^{2(m-1)} \sum_{\gamma=0}^{m-1} t_{\kappa} \, c_{\gamma} \, f(\kappa + \gamma).
\end{equation}

After reviewing the definitions of Gausslets and overlap integrals in Sec.~\ref{sec:background}, the effect of a uniform orbital rescaling in Sec.~\ref{sec:rescaling}, and overlap integral symmetries in Sec.~\ref{sec:symmetries}, we present algorithms and pseudocode for efficient evaluation of these integrals in Sec.~\ref{sec:algorithms}, followed by numerical experiments. A reference implementation is available as an open source repository at \cite{gausslectronitegrate}.

\section{Background theory and notation}
\label{sec:background}

The section summarizes the definition of Gausslets on a cubic lattice and the overlap integrals relevant to electronic-structure calculations. Throughout this work, $\norm{\cdot}$ denotes the Euclidean norm.

\subsection{Gausslet orbital basis}

We summarize the definition of Gausslets introduced in \cite{White2017}. A Gausslet in one dimension is defined as:
\begin{equation}
\label{eq:gausslet_def}
G(x) = \sum_{\alpha=-L}^L b_{\alpha} \, g_{\alpha}(x)
\end{equation}
where $\{b_{\alpha}\} \in \R^{2 L + 1}$ are precomputed coefficients and $g_{\alpha}$ for $\alpha \in \Z$ is an elementary Gaussian function defined on a grid with spacing $\frac{1}{3}$ and fixed width $\frac{1}{3}$ (cf.~\cite[(A1)]{White2017}):
\begin{equation}
\label{eq:elementary_gaussian_def}
g_{\alpha}(x) \coloneq \e^{-\frac{1}{2} \left(\frac{x - \alpha/3}{1/3}\right)^2} = \e^{-\frac{1}{2} (3 x - \alpha)^2}, \quad x \in \R.
\end{equation}

The Gausslets are orthonormalized such that
\begin{equation}
\label{eq:gausslet_orthonormalization}
\int_{\R} G(x) \, G(x - i) \, \ud x = \delta_{i,0} \quad \text{for all } i \in \Z.
\end{equation}

Taking Cartesian outer products in three dimensions and shifting by integer offsets defines an orbital basis $\{ \mathcal{G}_i \}_{i \in \Lambda}$ on the cubic lattice $\Lambda \coloneq \Z^3$:
\begin{equation}
\label{eq:gausslet_orbitals_def}
\mathcal{G}_i(r) \coloneq G(r_1 - i_1) \, G(r_2 - i_2) \, G(r_3 - i_3)
\end{equation}
for all $r \in \R^3$ and $i \in \Lambda$. Their orthonormalization follows from the one-dimensional analog in Eq.~\eqref{eq:gausslet_orthonormalization}.

\subsection{Overlap integrals for molecular Hamiltonians}

The molecular Hamiltonian governing electronic structure calculations comprises terms describing the kinetic energy of the electrons, the Coulomb attraction between the electrons and the atomic nuclei, and the mutual Coulomb repulsion between the electrons \cite{Helgaker2000}. We use atomic units and the Born-Oppenheimer approximation (fixed nuclei).

The following overlap integrals are expressed in terms of a (generic) smooth, real-valued spatial orbital basis $\{ \phi_i \}$ with $\phi_i \in L^2(\R^3, \R)$ for all $i$, adhering to the orthonormalization property
\begin{equation}
\label{eq:orbital_orthonormalization}
\int_{\R^3} \phi_i(r) \, \phi_j(r) \, \ud^3 r = \delta_{i,j} \quad \text{for all } i, j.
\end{equation}
Subsequently, we will specialize to Gausslets, i.e., $\phi_i = \mathcal{G}_i$ for all $i$.

The kinetic overlap integrals are defined as
\begin{equation}
\label{eq:tkin3_ij_def}
\mathbf{k}_{i,j} \coloneq -\frac{1}{2} \int_{\R^3} \phi_i(r) \, \Delta \phi_j(r) \, \ud^3 r,
\end{equation}
where $\Delta = \partial_{r_1}^2 + \partial_{r_2}^2 + \partial_{r_3}^2$ is the Laplace operator in three dimensions. Using integration by parts and assuming that the orbital functions vanish sufficiently fast as $\norm{r} \to \infty$ (such that boundary terms are absent), we obtain the more symmetric representation
\begin{equation}
\label{eq:tkin3_ij}
\begin{split}
\mathbf{k}_{i,j} &= \frac{1}{2} \int_{\R^3} \left(\nabla \phi_i(r)\right) \cdot \left(\nabla \phi_j(r)\right) \ud^3 r\\
&= \frac{1}{2} \sum_{n=1}^3 \int_{\R^3} \left(\partial_{r_n} \phi_i(r)\right) \left(\partial_{r_n} \phi_j(r)\right) \ud^3 r,
\end{split}
\end{equation}
where $\nabla$ is the nabla operator.

The Coulomb attraction exerted by an atomic nucleus at position $R \in \R^3$ with charge $Z \in \N$ is represented by the overlap integral
\begin{equation}
\label{eq:nuclear_ij_def}
\mathfrak{n}_{i,j}^Z(R) \coloneq \int_{\R^3} \phi_i(r) \, \frac{Z}{\norm{r - R}} \phi_j(r) \, \ud^3 r.
\end{equation}
The attraction by several nuclei is the sum over their respective overlap integrals.

The Hamiltonian terms describing the Coulomb repulsion between electrons are captured by the electron repulsion integral (ERI) tensor with entries:
\begin{equation}
\label{eq:coul_ijkl_def}
\begin{split}
&\coul{ij \vert k\ell} \coloneq \\
&\int_{\R^3} \int_{\R^3} \phi_i(r) \phi_j(r) \frac{1}{\norm{r - r'}} \phi_k(r') \phi_{\ell}(r') \, \ud^3 r \, \ud^3 r'.
\end{split}
\end{equation}
Here, $\norm{r - r'}$ is the Euclidean distance between the points $r$ and $r'$ in three dimensions.

When using Gausslets as orbital basis, the electron repulsion integral tensor can be compactified based on the ``integral diagonal approximation'' (IDA) due to the polynomial completeness and moment property of the Gausslets \cite{White2017}:
\begin{equation}
\coul{ij \vert k\ell} \approx \delta_{ij} \delta_{k\ell} \, \mathbf{v}_{i,k}
\end{equation}
with
\begin{equation}
\label{eq:v_ij_def}
\mathbf{v}_{i,j} \coloneq \frac{1}{m_i m_j} \int_{\R^3} \int_{\R^3} \frac{\phi_i(r) \phi_j(r')}{\norm{r - r'}} \, \ud^3 r \, \ud^3 r'
\end{equation}
and
\begin{equation}
\label{eq:m_i_def}
m_i \coloneq \int_{\R^3} \phi_i(r) \, \ud^3 r.
\end{equation}
We will denote the $\mathbf{v}$ matrix as ERIDA.

\section{Rescaling the orbital basis}
\label{sec:rescaling}

We study the effect of a global rescaling of the orbital basis by a fixed factor $s > 0$. Specifically, the rescaled orbital functions are defined as
\begin{equation}
\tilde{\phi}_i(r) \coloneq s^{3/2} \phi_i(s r) \quad \text{for all } r \in \R^3 \text{ and } i.
\end{equation}
By this convention, $s > 1$ corresponds to a finer resolution.

Orbital orthonormalization is retained since
\begin{multline}
\int_{\R^3} \tilde{\phi}_i(r) \, \tilde{\phi}_j(r) \, \ud^3 r = \int_{\R^3} \phi_i(s r) \, \phi_j(s r) \, s^3 \, \ud^3 r\\
= \int_{\R^3} \phi_i(r') \, \phi_j(r') \, \ud^3 r' \stackrel{\eqref{eq:orbital_orthonormalization}}{=} \delta_{i,j} \quad \text{for all } i, j.
\end{multline}

The global rescaling transforms the kinetic overlap integrals as
\begin{equation}
\begin{split}
\tilde{\mathbf{k}}_{i,j}%
&= -\frac{1}{2} \int_{\R^3} \tilde{\phi}_i(r) \, \Delta_r \tilde{\phi}_j(r) \, \ud^3 r\\
&= -\frac{1}{2} \int_{\R^3} \phi_i(s r) \, \Delta_r \phi_j(s r) \, s^3 \, \ud^3 r\\
&= -\frac{1}{2} \int_{\R^3} \phi_i(r') \, s^2 \Delta_{r'} \phi_j(r') \, \ud^3 r'\\
&= s^2 \, \mathbf{k}_{i, j} \quad \text{for all } i, j.
\end{split}
\end{equation}
Here, the notation $\Delta_r$ indicates that the Laplace operator is to be evaluated with respect to the $r$ variable. The factor $s^2$ originates from taking second derivatives and the chain rule.

The nuclear repulsion integrals are transformed as
\begin{equation}
\begin{split}
\tilde{\mathfrak{n}}_{i,j}^Z(R)%
&= \int_{\R^3} \tilde{\phi}_i(r) \, \frac{Z}{\norm{r - R}} \tilde{\phi}_j(r) \, \ud^3 r\\
&= \int_{\R^3} \phi_i(s r) \, \frac{s Z}{\norm{s r - s R}} \phi_j(s r) \, s^3 \, \ud^3 r\\
&= \int_{\R^3} \phi_i(r') \, \frac{s Z}{\norm{r' - s R}} \phi_j(r') \, \ud^3 r'\\
&= s \, \mathfrak{n}_{i, j}^Z(s R) \quad \text{for all } R \in \R^3 \text{ and } i, j.
\end{split}
\end{equation}

Similarly, the electronic Coulomb overlap integrals transform under the rescaling as
\begin{equation}
\begin{split}
&\coul{ij \vert k\ell}\underaccent{\tilde}{}%
= \int_{\R^3} \int_{\R^3} \tilde{\phi}_i(r) \tilde{\phi}_j(r) \frac{1}{\norm{r - r'}} \tilde{\phi}_k(r') \tilde{\phi}_{\ell}(r') \, \ud^3 r \, \ud^3 r'\\
&= \int_{\R^3} \int_{\R^3} \phi_i(s r) \phi_j(s r) \frac{s}{\norm{s r - s r'}}\\
&\hspace{0.4\linewidth} \times \phi_k(s r') \phi_{\ell}(s r') \, s^6 \, \ud^3 r \, \ud^3 r'\\
&= \int_{\R^3} \int_{\R^3} \phi_i(r) \phi_j(r) \frac{s}{\norm{r - r'}} \phi_k(r') \phi_{\ell}(r') \, \ud^3 r \, \ud^3 r'\\
&= s \, \coul{ij \vert k\ell} \quad \text{for all } i, j, k, \ell.
\end{split}
\end{equation}

Under the integral diagonal approximation, one has to additionally take the rescaled weight factor in Eq.~\eqref{eq:m_i_def} into account: for all $i$,
\begin{equation}
\begin{split}
\tilde{m}_i%
&= \int_{\R^3} \tilde{\phi}_i(r) \, \ud^3 r%
 = \int_{\R^3} s^{3/2} \phi_i(s r) \, \ud^3 r\\
&= s^{-3/2} \int_{\R^3} \phi_i(r') \, \ud^3 r'%
 = s^{-3/2} \, m_i.
\end{split}
\end{equation}
Thus
\begin{equation}
\begin{split}
\tilde{\mathbf{v}}_{i,j}%
&= \frac{1}{\tilde{m}_i \tilde{m}_j} \int_{\R^3} \int_{\R^3} \frac{\tilde{\phi}_i(r) \tilde{\phi}_j(r')}{\norm{r - r'}} \, \ud^3 r \, \ud^3 r'\\
&= \frac{s^3}{m_i m_j} \int_{\R^3} \int_{\R^3} \frac{\phi_i(s r) \phi_j(s r')}{\norm{s r - s r'}} \, s^4 \, \ud^3 r \, \ud^3 r'\\
&= s \, \mathbf{v}_{i, j} \quad \text{for all } i, j.
\end{split}
\end{equation}

\section{Symmetries}
\label{sec:symmetries}

The overlap integrals have various symmetries, which we describe in this section. We will use them to avoid redundant calculations. From now on, we will employ Gausslets $\mathcal{G}_i$ as orbital basis, indexed by points $i \in \Lambda$, where $\Lambda \coloneq \Z^3$ is the cubic lattice.

In anticipation of exploiting translational and octahedral symmetries, we will use the finite grid (for odd integer $d$)
\begin{equation}
\Lambda_d \coloneq \left\{ i \in \Lambda : \norm{i}_{\infty} \le \frac{1}{2} (d - 1) \right\}
\end{equation}
in our framework, where $\norm{i}_{\infty} = \max(\abs{i_1}, \abs{i_2}, \abs{i_3})$ denotes the maximum norm. By construction, $\Lambda_d$ contains $d^3$ points.

\subsection{Translations}
\label{sec:translations}

A simultaneous translation of the lattice points defining an overlap integral preserves the kinetic, ERI, and ERIDA integral values. For example, for arbitrary $i, j \in \Lambda$ and any translation vector $\ell \in \Lambda$, $\mathbf{k}_{i,j} = \mathbf{k}_{i-\ell,j-\ell}$, and analogously for the ERI tensor defined in Eq.~\eqref{eq:coul_ijkl_def} and the ERIDA matrix $\mathbf{v}$ in Eq.~\eqref{eq:v_ij_def}. Note, however, that the translational invariance does not hold for the nuclear attraction integral in Eq.~\eqref{eq:nuclear_ij_def} when regarding the nuclear position as fixed.

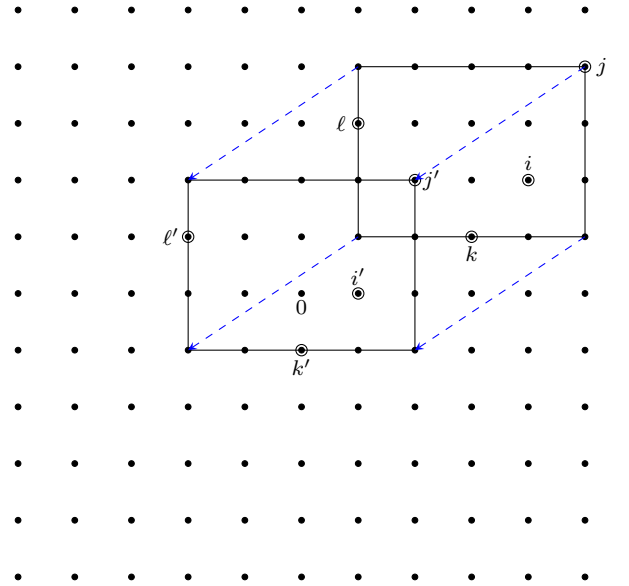
\begin{figure}[!ht]
\centering
\input{translation.tikz}
\caption{Exploiting translational symmetry of an ERI term $\coul{ij \vert k\ell}$ (shown for a two-dimensional grid for visual clarity). The bounding box of the four points is translated such that its center is as close as possible to the origin. Only the integral of the translated points needs to be evaluated.}
\label{fig:translation}
\end{figure}

A practical question is how to choose representative lattice points for each type of overlap integrals. We use the approach visualized in Fig.~\ref{fig:translation} for the ERI tensor $\coul{ij \vert k\ell} $. The three-dimensional bounding box of the four grid points is shifted towards the origin (by integer translations) such that its center is as close as possible to the origin. Specifically, each coordinate of the bounding box center is in $\{ -\frac{1}{2}, 0, \frac{1}{2} \}$. The resolution of the $\pm \frac{1}{2}$ ambiguity will be described below. The points $i, j, k, \ell$ are shifted accordingly. Note that the shifted points will remain elements of the finite grid $\Lambda_d$ if $i, j, k, \ell \in \Lambda_d$.

We use an analogous procedure for the ERIDA integrals $\mathbf{v}_{i,j}$ with $i, j \in \Lambda$.

\subsection{Permutations}
\label{sec:permutations}

The overlap integrals are symmetric with respect to the following index permutations: for all $i, j \in \Lambda$,
\begin{equation}
\mathbf{k}_{i,j} = \mathbf{k}_{j,i}, \quad \mathfrak{n}_{i,j}^Z(R) = \mathfrak{n}_{j,i}^Z(R), \quad \mathbf{v}_{i,j} = \mathbf{v}_{j,i}.
\end{equation}
These symmetries follow directly from the integral definitions.

One further observes that the ERI tensor $\coul{ij \vert k\ell} $ is invariant when interchanging $i \leftrightarrow j$, $k \leftrightarrow \ell$, or $(i, j) \leftrightarrow (k, \ell)$ for any $i, j, k, \ell \in \Lambda$, as well as any combination of these. Explicitly,
\begin{equation}
\begin{split}
&\hspace{4mm} \coul{ij \vert k\ell} = \coul{ji \vert k\ell} = \coul{ij \vert \ell k} = \coul{ji \vert \ell k} \\
&= \coul{k\ell \vert ij} = \coul{k\ell \vert ji} = \coul{\ell k \vert ij} = \coul{\ell k \vert ji}.
\end{split}
\end{equation}
From a group theory perspective, these permutations correspond to the dihedral group $\Dfour$ of order 8 (symmetry group of a square) when arranging the indices on the corners of the square as
\begin{equation*}
\input{eri_dihedral_symmetry.tikz}
\end{equation*}

\subsection{Octahedral symmetry}
\label{sec:octahedral}

The crystallographic point group of the cubic lattice $\Lambda = \Z^3$ is $\Oh$ \cite{Tinkham2003}, consisting of all rotations and reflections which map the lattice points to themselves while leaving the origin invariant. For Gausslet orbitals as defined in Eq.~\eqref{eq:gausslet_orbitals_def}, the kinetic, ERI, and ERIDA overlap integrals inherit this symmetry: that is, the integral value remains unchanged when acting with an $\Oh$ group element on the lattice points defining the integral. Note that this invariance requires the factorized form of the Gausslet orbitals with respect to Cartesian coordinates.

\section{Algorithmic evaluation of overlap integrals}
\label{sec:algorithms}

This section develops efficient algorithms for evaluating overlap integrals using the Gausslet orbitals $\{ \mathcal{G}_i \}$.

\subsection{Kinetic overlap integrals}

The integral in Eq.~\eqref{eq:tkin3_ij} factorizes with respect to Cartesian coordinate directions by definition of the Gausslets, and it thus suffices to consider one direction, say the $x$-direction.

The derivative of an elementary Gaussian, Eq.~\eqref{eq:elementary_gaussian_def}, is
\begin{equation}
g_{\alpha}'(x) = -3 (3 x - \alpha) \e^{-\frac{1}{2} (3 x - \alpha)^2}.
\end{equation}
Computing the corresponding overlap integral (for $\alpha, \beta \in \Z$) leads to
\begin{align}
\int_{\R} g_{\alpha}'(x) \, g_{\beta}'(x) \, \ud x &= T(\alpha - \beta),\\
T(\ell) &\coloneq \frac{3 \sqrt{\pi}}{2} \left(1 - \frac{1}{2} \ell^2\right) \e^{-\frac{1}{4} \ell^2}.
\end{align}
With that, we can now evaluate (for $\ell \in \Z$)
\begin{equation}
\begin{split}
k_{\ell} &\coloneq \frac{1}{2} \int_{\R} G'(x - \ell) \, G'(x) \, \ud x\\
&= \frac{1}{2} \sum_{\alpha,\beta=-L}^L b_{\alpha} b_{\beta} \int_{\R} g_{\alpha}'(x - \ell) \, g_{\beta}'(x) \, \ud x\\
&= \frac{1}{2} \sum_{\alpha,\beta=-L}^L b_{\alpha} b_{\beta} \, T\big(3 \ell + \alpha - \beta\big).
\end{split}
\end{equation}

We obtain the following expression for the kinetic overlap integral in three dimensions (with $i, j \in \Lambda$):
\begin{equation}
\begin{split}
\mathbf{k}_{i,j}%
&= k_{i_1 - j_1} \delta_{i_2, j_2} \delta_{i_3, j_3}\\
&+ \delta_{i_1,j_1} k_{i_2 - j_2} \delta_{i_3, j_3}\\
&+ \delta_{i_1,j_1} \delta_{i_2, j_2} k_{i_3 - j_3}.
\end{split}
\end{equation}
The delta functions result from the orthonormalization of the one-dimensional Gausslets (see Eq.~\eqref{eq:gausslet_orthonormalization}) and the fact that the derivatives in Eq.~\eqref{eq:tkin3_ij} act on one coordinate direction at a time.

\subsection{Nuclear overlap integrals}

To evaluate the nuclear overlap integrals in Eq.~\eqref{eq:nuclear_ij_def} for the $\{ \mathcal{G}_i \}$ orbital basis, we start from the known analytic expression for the Coulomb overlap integral of a (normalized) Gaussian function and a point charge \cite{Boys1950}. We use the following convention for a normalized Gaussian function with width $w > 0$:
\begin{equation}
\label{eq:generic_gaussian_function_def}
f_w(r) \coloneq \frac{1}{(\sqrt{\pi} w)^3} \e^{-\norm{r / w}^2}, \quad r \in \R^3.
\end{equation}
The overlap integral for a Gaussian centered at the origin and a point charge located at $R \in \R^3$ is defined as
\begin{equation}
\label{eq:nuclear_integral_def}
K_w(\norm{R}) \coloneqq \int_{\R^3} \frac{f_w(r)}{\norm{r - R}} \, \ud^3 r.
\end{equation}
Due to translation and rotation invariance, the integral value only depends on the Euclidean length $\norm{R}$. The analytic expression (for $x = \norm{R} > 0$) reads \cite{Boys1950}
\begin{equation}
K_w(x) = \frac{1}{x} \erf\!\left(\frac{x}{w}\right),
\end{equation}
where $\erf(\cdot)$ is the error function. We can extend $K_w$ by taking the limit $x \to 0$:
\begin{equation}
K_w(0) = \lim_{x \to 0} K_w(x) = \frac{2}{\sqrt{\pi} w}.
\end{equation}
Substituting the definition \eqref{eq:gausslet_orbitals_def} together with the expansion \eqref{eq:gausslet_def} into Eq.~\eqref{eq:nuclear_ij_def} leads to
\begin{multline}
\mathfrak{n}_{i,j}^Z(R) = \int_{\R^3} \prod_{n=1}^3 \Bigg(\sum_{\alpha_n,\beta_n=-L}^L b_{\alpha_n} b_{\beta_n} \, g_{\alpha_n}(r_n - i_n)\\
\times g_{\beta_n}(r_n - j_n) \Bigg) \frac{Z}{\norm{r - R}} \, \ud^3 r.
\end{multline}
We now insert the definition \eqref{eq:elementary_gaussian_def} to evaluate products of two elementary Gaussians:
\begin{multline}
\label{eq:elementary_gaussian_product}
g_{\alpha_n}(r_n - i_n) \, g_{\beta_n}(r_n - j_n)\\
= \e^{-\left( 3 (r_n - \frac{i_n + j_n}{2}) - \frac{\alpha_n + \beta_n}{2}\right)^2} \e^{-\left(\frac{3 (i_n - j_n) + (\alpha_n - \beta_n)}{2}\right)^2}.
\end{multline}
The first factor is a Gaussian, and the second factor is independent of $r$, so we can move it out of the integral. Together with the definition \eqref{eq:nuclear_integral_def}, one obtains
\begin{equation}
\begin{split}
&\mathfrak{n}_{i,j}^Z(R) = Z \left(\frac{\sqrt{\pi}}{3}\right)^3 \times\\
&\sum_{\alpha_1,\beta_1=-L}^L b_{\alpha_1} b_{\beta_1} \e^{-\left(\frac{3 (i_1 - j_1) + (\alpha_1 - \beta_1)}{2}\right)^2}\\
&\sum_{\alpha_2,\beta_2=-L}^L b_{\alpha_2} b_{\beta_2} \e^{-\left(\frac{3 (i_2 - j_2) + (\alpha_2 - \beta_2)}{2}\right)^2}\\
&\sum_{\alpha_3,\beta_3=-L}^L b_{\alpha_3} b_{\beta_3} \e^{-\left(\frac{3 (i_3 - j_3) + (\alpha_3 - \beta_3)}{2}\right)^2}\\
&\times K_{\frac{1}{3}}\left(\norm*{\begin{pmatrix}\frac{i_n + j_n}{2} + \frac{\alpha_n + \beta_n}{6} - R_n\end{pmatrix}_{n=1,2,3}}\right).
\end{split}
\end{equation}
There are $(2 L + 1)^6$ summands, so a literal evaluation of this expression is challenging for realistic values $L \approx 100$. We circumvent this problem by a reorganization of the summation, noting that $K_{\frac{1}{3}}$ only depends on the summation indices via $\alpha_n + \beta_n$. Let
\begin{equation}
\label{eq:gausslet_factors}
c_{\Delta i, \mu} \coloneq \frac{\sqrt{\pi}}{3} \sum_{\alpha,\beta=-L}^L b_{\alpha} b_{\beta} \e^{-\left(\frac{3 \Delta i + (\alpha - \beta)}{2}\right)^2} \delta_{\alpha + \beta, \mu}
\end{equation}
for $\Delta i \in \Z$ and $\mu \in \{-2 L, \dots, 2 L\}$. Then
\begin{multline}
\mathfrak{n}_{i,j}^Z(R) = Z \sum_{\mu_1,\mu_2,\mu_3=-2L}^{2L} c_{i_1 - j_1, \mu_1} \, c_{i_2 - j_2, \mu_2} \, c_{i_3 - j_3, \mu_3}\\
\times K_{\frac{1}{3}}\left(\norm*{\begin{pmatrix}\frac{i_n + j_n}{2} + \frac{\mu_n}{6} - R_n\end{pmatrix}_{n=1,2,3}}\right).
\end{multline}
This expression contains $(4 L + 1)^3$ summands and is amenable to a direct evaluation. We can further reduce the computational cost by filtering out coefficients $\abs{c_{\Delta i, \mu}} \le \epsilon_{\text{tol}}$, where $\epsilon_{\text{tol}}$ is a user-defined truncation tolerance.

\subsection{Electron repulsion integrals}

To efficiently evaluate the overlap integrals in Eq.~\eqref{eq:coul_ijkl_def} for the $\{ \mathcal{G}_i \}$ basis, we use the known analytic expression for the Coulomb overlap integral of two (normalized) Gaussian functions \cite{Boys1950}. We use the same convention for a normalized Gaussian function $f_w$ with width $w > 0$ as before in Eq.~\eqref{eq:generic_gaussian_function_def}. The Coulomb integral of two Gaussians centered at $p, q \in \R^3$ is defined as
\begin{equation}
\label{eq:coulomb_integral_def}
C_w(\norm{p - q}) \coloneqq \int_{\R^3} \int_{\R^3} f_w(r - p) \frac{1}{\norm{r - r'}} f_w(r' - q) \, \ud^3 r \, \ud^3 r'.
\end{equation}
Due to translation and rotation invariance, the integral value only depends on the distance $\norm{p - q}$. The analytic expression (for $p \neq q$, i.e., $x = \norm{p - q} > 0$) reads \cite{Boys1950}
\begin{equation}
\label{eq:coulomb_integral_formula}
C_w(x) = \frac{1}{x} \erf\!\left(\frac{1}{\sqrt{2} w} x\right),
\end{equation}
where $\erf(\cdot)$ is the error function. We can extend $C_w$ by taking the limit $x \to 0$:
\begin{equation}
C_w(0) = \lim_{x \to 0} C_w(x) = \sqrt{\frac{2}{\pi}} \frac{1}{w}.
\end{equation}
Substituting the definition \eqref{eq:gausslet_orbitals_def} together with the expansion \eqref{eq:gausslet_def} into Eq.~\eqref{eq:coul_ijkl_def} leads to
\begin{multline}
\coul{ij \vert k\ell} = \int_{\R^3} \int_{\R^3}\\
\prod_{n=1}^3 \Bigg(\sum_{\alpha_n,\beta_n,\gamma_n,\delta_n=-L}^L b_{\alpha_n} b_{\beta_n} b_{\gamma_n} b_{\delta_n} \, g_{\alpha_n}(r_n - i_n)\\
g_{\beta_n}(r_n - j_n) \, g_{\gamma_n}(r'_n - k_n) \, g_{\delta_n}(r'_n - \ell_n) \Bigg)\\
\times \frac{1}{\norm{r - r'}} \, \ud^3 r \, \ud^3 r'.
\end{multline}
We evaluate products of two elementary Gaussians as before in Eq.~\eqref{eq:elementary_gaussian_product}. Together with the definition \eqref{eq:coulomb_integral_def}, one obtains
\begin{equation}
\begin{split}
&\coul{ij \vert k\ell} = \left(\frac{\sqrt{\pi}}{3}\right)^6 \times\\
&\sum_{\alpha_1,\beta_1,\gamma_1,\delta_1=-L}^L b_{\alpha_1} b_{\beta_1} b_{\gamma_1} b_{\delta_1}\\
&\quad \times \e^{-\left(\frac{3 (i_1 - j_1) + (\alpha_1 - \beta_1)}{2}\right)^2 - \left(\frac{3 (k_1 - \ell_1) + (\gamma_1 - \delta_1)}{2}\right)^2}\\
&\sum_{\alpha_2,\beta_2,\gamma_2,\delta_2=-L}^L b_{\alpha_2} b_{\beta_2} b_{\gamma_2} b_{\delta_2}\\
&\quad \times \e^{-\left(\frac{3 (i_2 - j_2) + (\alpha_2 - \beta_2)}{2}\right)^2 - \left(\frac{3 (k_2 - \ell_2) + (\gamma_2 - \delta_2)}{2}\right)^2}\\
&\sum_{\alpha_3,\beta_3,\gamma_3,\delta_3=-L}^L b_{\alpha_3} b_{\beta_3} b_{\gamma_3} b_{\delta_3}\\
&\quad \times \e^{-\left(\frac{3 (i_3 - j_3) + (\alpha_3 - \beta_3)}{2}\right)^2 - \left(\frac{3 (k_3 - \ell_3) + (\gamma_3 - \delta_3)}{2}\right)^2}\\
&\times C_{\frac{1}{3}}\left(\norm*{\begin{pmatrix}\frac{i_n + j_n}{2} - \frac{k_n + \ell_n}{2} + \frac{\alpha_n + \beta_n}{6} - \frac{\gamma_n + \delta_n}{6}\end{pmatrix}_{n=1,2,3}}\right).
\end{split}
\end{equation}
There are $(2 L + 1)^{12}$ summands, so a literal evaluation of this expression is infeasible for realistic values $L \approx 100$. As for the nuclear overlap integrals, we circumvent this problem by a reorganization of the summation, noting that $C_{\frac{1}{3}}$ only depends on the summation indices via $(\alpha_n + \beta_n) - (\gamma_n + \delta_n)$. Let $c_{\Delta i, \mu}$ as introduced in Eq.~\eqref{eq:gausslet_factors} for $\Delta i \in \Z$ and $\mu \in \{-2 L, \dots, 2 L\}$, and define
\begin{equation}
\label{eq:eri_gausslet_factor_products_def}
d_{\Delta i, \Delta k, \xi} \coloneq \sum_{\mu,\nu=-2L}^{2L} c_{\Delta i, \mu} c_{\Delta k, \nu} \, \delta_{\mu - \nu, \xi}
\end{equation}
for $\Delta i, \Delta k \in \Z$ and $\xi \in \{-4 L, \dots, 4 L\}$. Then
\begin{equation}
\label{eq:coul_ijkl_triple_sum}
\begin{split}
\coul{ij \vert k\ell} = %
&\sum_{\xi_1=-4L}^{4L} d_{i_1 - j_1, k_1 - \ell_1, \xi_1}\\
&\sum_{\xi_2=-4L}^{4L} d_{i_2 - j_2, k_2 - \ell_2, \xi_2}\\
&\sum_{\xi_3=-4L}^{4L} d_{i_3 - j_3, k_3 - \ell_3, \xi_3}\\
&\times C_{\frac{1}{3}}\left(\norm*{\begin{pmatrix}\frac{i_n + j_n}{2} - \frac{k_n + \ell_n}{2} + \frac{\xi_n}{6}\end{pmatrix}_{n=1,2,3}}\right).
\end{split}
\end{equation}
To further reorganize the triple sum and to precompute a look-up table for the Coulomb integral function $C_{\frac{1}{3}}$, we notice that the argument of $C_{\frac{1}{3}}$ can be mapped to an integer by multiplying the inner vector by $6$ and squaring its norm:
\begin{multline}
C_{\frac{1}{3}}\left(\norm*{\begin{pmatrix}\frac{i_n + j_n}{2} - \frac{k_n + \ell_n}{2} + \frac{\xi_n}{6}\end{pmatrix}_{n=1,2,3}}\right) \\
= \tilde{C}_{\frac{1}{3}}\left(\sum_{n=1}^3 \big(3 (i_n + j_n - k_n - \ell_n) + \xi_n\big)^2\right)
\end{multline}
with
\begin{equation}
\tilde{C}_{\frac{1}{3}}(x) \coloneq C_{\frac{1}{3}}(\sqrt{x} / 6), \quad x \in \R_{\ge 0}.
\end{equation}
This allows us to reorganize the summations in Eq.~\eqref{eq:coul_ijkl_triple_sum}: introduce (for $t_1, t_2, \Delta i_1, \Delta k_1, \Delta i_2, \Delta k_2 \in \Z$ and $\rho \in \N_0$)
\begin{multline}
\label{eq:eri_circular_contour_products_def}
f^{t_1, t_2}_{\Delta i_1, \Delta k_1, \Delta i_2, \Delta k_2, \rho} \coloneq \sum_{\xi_1,\xi_2=-4L}^{4L} d_{\Delta i_1, \Delta k_1, \xi_1} d_{\Delta i_2, \Delta k_2, \xi_2} \\
\times \delta_{(t_1 + \xi_1)^2 + (t_2 + \xi_2)^2, \rho},
\end{multline}
then
\begin{multline}
\label{eq:coul_ijkl_computation}
\coul{ij \vert k\ell} = %
\sum_{\rho=0}^{\rho_{\max}} f^{t_1, t_2}_{i_1 - j_1, k_1 - \ell_1, i_2 - j_2, k_2 - \ell_2, \rho}\\
\sum_{\xi_3=-4L}^{4L} d_{i_3 - j_3, k_3 - \ell_3, \xi_3} \, \tilde{C}_{\frac{1}{3}}\left(\rho + (t_3 + \xi_3)^2\right)
\end{multline}
with $\rho_{\max} = (\abs{t_1} + 4 L)^2 + (\abs{t_2} + 4 L)^2$ and $t_n = 3 (i_n + j_n - k_n - \ell_n)$ for $n = 1, 2, 3$. In practice, we evaluate $f^{t_1, t_2}_{\Delta i_1, \Delta k_1, \Delta i_2, \Delta k_2, \rho}$ on the fly for each tuple $(i, j, k, \ell)$.

We can further reduce the computational cost by filtering out coefficients $\abs{d_{\Delta i, \Delta k, \xi}} \le \epsilon_{\text{tol}}$, where the truncation threshold $\epsilon_{\text{tol}}$ is user-defined.

\medskip

We exploit symmetries in our implementation by first listing a minimal representative set of grid index tuples $(i, j, k, \ell) \in \Lambda_d^4$ such that all remaining ERI integrals (on $\Lambda_d$) can be obtained from this set by symmetry transformations of the grid indices. More formally, we introduce equivalence classes (for $i, j, k, \ell, i', j', k', \ell' \in \Lambda_d$) by the equivalence relation defined as
\begin{equation}
(i, j, k, \ell) \sim (i', j', k', \ell')
\end{equation}
precisely if $(i, j, k, \ell)$ can be mapped to $(i', j', k', \ell')$ by a translation, index permutation, and/or octahedral point group transformation as described in Sec.~\ref{sec:symmetries}. To select a representative tuple from each class, we will assign a unique lexicographical index to each tuple as follows. First, we enumerate individual grid points lexicographically: for $\ell \in \Lambda_d$, let
\begin{equation}
\gridindex{d}(\ell) \coloneq (\ell_1 + s_d) d^2 + (\ell_2 + s_d) d + (\ell_3 + s_d)
\end{equation}
with shift $s_d \coloneq \frac{1}{2} (d - 1)$. By construction, $\gridindex{d}(\ell) \in \{ 0, \dots, d^3 - 1 \}$. Analogously, we compute the lexicographical index of a tuple $(i, j, k, \ell) \in \Lambda_d^4$:
\begin{multline}
\eriindex{d}(i, j, k, \ell) \coloneq \\
\gridindex{d}(i) \, \abs{\Lambda_d}^3 + \gridindex{d}(j) \, \abs{\Lambda_d}^2 + \gridindex{d}(k) \, \abs{\Lambda_d} + \gridindex{d}(\ell)
\end{multline}
where $\abs{\Lambda_d} = d^3$ denotes the number of grid points.

We select a representative from each equivalence class as follows. Given $(i, j, k, \ell) \in \Lambda_d^4$, shift the points towards the origin based on their bounding box, as described in Sec.~\ref{sec:translations}, enumerating both $\pm \frac{1}{2}$ cases of each new bounding box coordinate (if the coordinate is not $0$). More formally, we collect the translated points in the set
\begin{multline}
\boxtrans(i, j, k, \ell) \coloneq \bigg\{ (i', j', k', \ell') \in \Lambda_d^4 \,\Big\vert\, \\
(i', j', k', \ell') = (i - s, j - s, k - s, \ell - s) \text{ for } s \in \Lambda, \\
\tfrac{1}{2} \big(\min(i'_n, j'_n, k'_n, \ell'_n) + \max(i'_n, j'_n, k'_n, \ell'_n)\big) \\
\in \{ -\tfrac{1}{2}, 0, \tfrac{1}{2} \} \text{ for } n = 1, 2, 3 \bigg\}.
\end{multline}
Next, we select the tuple with minimal lexicographical index that emerges from $\boxtrans(i, j, k, \ell)$ by index permutations and octahedral symmetry (Sec.~\ref{sec:permutations} and \ref{sec:octahedral}), i.e.,
\begin{equation}
(i'', j'', k'', \ell'') = \argmin_{(\tilde{i}, \tilde{j}, \tilde{k}, \tilde{\ell}) \in \mathcal{S}(i, j, k, \ell)} \eriindex{d}\big(\tilde{i}, \tilde{j}, \tilde{k}, \tilde{\ell}\big)
\end{equation}
with
\begin{equation}
\mathcal{S}(i, j, k, \ell) \coloneq \bigcup_{\substack{(i', j', k', \ell') \in \\ \boxtrans(i, j, k, \ell)}} \orb(i', j', k', \ell')
\end{equation}
and the permutation and octahedral symmetry ``orbit''
\begin{multline}
\orb(i', j', k', \ell') \coloneq \Big\{ \text{perm}\left(u(i'), u(j'), u(k'), u(\ell')\right) : \\
\text{perm} \in \Dfour, u \in \Oh \Big\}.
\end{multline}
Here, the notation $u(i)$ denotes the action of a point group element $u \in \Oh$ on a lattice point $i$. The tuple $(i'', j'', k'', \ell'')$ is the representative instance of the initial tuple $(i, j, k, \ell)$.

\begin{table}
\begin{tabular}{|crrrl|}
\hline
$\Lambda_d$ dim. & $\abs{\Lambda_d}$ &       \# dense &   \# symm. & reduction \\
                 &                   &            ERI &        ERI &           \\
\hline
$1 \times 1 \times 1$    &         1 &              1 &          1 & 1         \\
$3 \times 3 \times 3$    &        27 &         531441 &       1027 & 0.0019    \\
$5 \times 5 \times 5$    &       125 &      244140625 &     138187 & 0.00057   \\
$7 \times 7 \times 7$    &       343 &    13841287201 &    3571977 & 0.00026   \\
$9 \times 9 \times 9$    &       729 &   282429536481 &   39277113 & 0.00014   \\
$11 \times 11 \times 11$ &      1331 &  3138428376721 &  261244251 & 0.000083  \\
$13 \times 13 \times 13$ &      2197 & 23298085122481 & 1250296187 & 0.000054  \\
\hline
\end{tabular}
\caption{Number of dense ERI tensor entries and symmetry reduction for increasing grid dimensions. The number of dense entries in the third column equals $\abs{\Lambda_d}^4$. The fourth column shows the number of integral evaluations actually required due to symmetry reduction, and the last column shows the corresponding reduction factor.}
\label{tab:eri_num_entries}
\end{table}

\begin{algorithm}[H]
\caption{Enumerate representative ERI indices}
\label{alg:enumerate_representative_eri}
\begin{algorithmic}
\Require $\Lambda_d$ for odd positive integer $d$
\State $s_d \gets \frac{1}{2} (d - 1)$
\State $\text{repr\_list} \gets \emptyset$
\For{$i_1, j_1, k_1, \ell_1 \in \left\{ -s_d, \dots, s_d \right\}$}
    \LineComment{$x$-coordinate of bounding box center}
    \State $b_1 \gets \frac{1}{2}(\min(i_1, j_1, k_1, \ell_1) + \max(i_1, j_1, k_1, \ell_1))$
    \If{$b_1 \notin \{-\frac{1}{2}, 0, \frac{1}{2}\}$}
        \State continue
    \EndIf
    \For{$i_2, j_2, k_2, \ell_2 \in \left\{ -s_d, \dots, s_d \right\}$}
        \LineComment{$y$-coordinate of bounding box center}
        \State $b_2 \gets \frac{1}{2}(\min(i_2, j_2, k_2, \ell_2) + \max(i_2, j_2, k_2, \ell_2))$
        \If{$b_2 \notin \{-\frac{1}{2}, 0, \frac{1}{2}\}$}
            \State continue
        \EndIf
        \For{$i_3, j_3, k_3, \ell_3 \in \left\{ -s_d, \dots, s_d \right\}$}
            \LineComment{$z$-coordinate of bounding box center}
            \State $b_3 \gets \frac{1}{2}(\min(i_3, j_3, k_3, \ell_3) + \max(i_3, j_3, k_3, \ell_3))$
            \If{$b_3 \notin \{-\frac{1}{2}, 0, \frac{1}{2}\}$}
                \State continue
            \EndIf
            \State $\text{idx} \gets \eriindex{d}(i, j, k, \ell)$
            \State is\_minimal\_index $\gets$ true
            \For{$\sigma_1, \sigma_2, \sigma_3 \in \{ 0, 1 \}$}
                \LineComment{translation bias}
                \State $\text{bias} \gets (2 \sigma_1 b_1, 2 \sigma_2 b_2, 2 \sigma_3 b_3)$
                 \If{not \begin{varwidth}[t]{0.6\linewidth}\textproc{is\_min\_orbit\_eri\_index}($\Lambda_d$, $(i - \text{bias}, j - \text{bias}, k - \text{bias}, \ell - \text{bias})$, idx)\end{varwidth}}
                    \State is\_minimal\_index $\gets$ false
                \EndIf
            \EndFor
            \If{is\_minimal\_index}
                \State add $(i, j, k, \ell)$ to $\text{repr\_list}$
            \EndIf
        \EndFor
    \EndFor
\EndFor
\State \Return repr\_list

\medskip

\Function{is\_min\_orbit\_eri\_index}{$\Lambda_d$, $(i, j, k, \ell)$, $\text{idx}$}
    \LineComment{iterate over the 48 group elements of $\Oh$}
    \For{$u \in \Oh$}
    \State $(i', j', k', \ell') \gets \left(u(i), u(j), u(k), u(\ell)\right)$
        \LineComment{dihedral permutation yielding minimal tuple index}
        \If{$\gridindex{d}(i') > \gridindex{d}(j')$}
            \State swap $i' \leftrightarrow j'$
        \EndIf
        \If{$\gridindex{d}(k') > \gridindex{d}(\ell')$}
            \State swap $k' \leftrightarrow \ell'$
        \EndIf
        \If{$\gridindex{d}(i') \, \abs{\Lambda_d} + \gridindex{d}(j') > \gridindex{d}(k') \, \abs{\Lambda_d} + \gridindex{d}(\ell')$}
            \State swap $(i', j') \leftrightarrow (k', \ell')$
        \EndIf
        \If{$\eriindex{d}(i', j', k', \ell') < \text{idx}$}
            \State \Return false
        \EndIf
    \EndFor
    \State \Return true
\EndFunction
\end{algorithmic}
\end{algorithm}

Algorithm~\ref{alg:enumerate_representative_eri} implements the described enumeration of representative ERI indices based on symmetries as pseudocode. The actual evaluation of ERI integrals will be performed by another dedicated function (see below).

\begin{figure}[!ht]
\centering
\subfloat[number of ERI tensor entries]{\includegraphics[width=0.9\columnwidth]{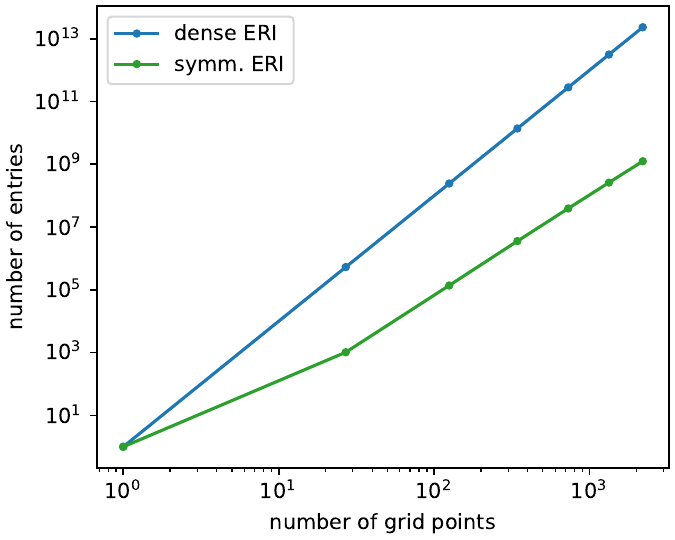}}\\
\subfloat[reduction factor due to symmetries]{\includegraphics[width=0.9\columnwidth]{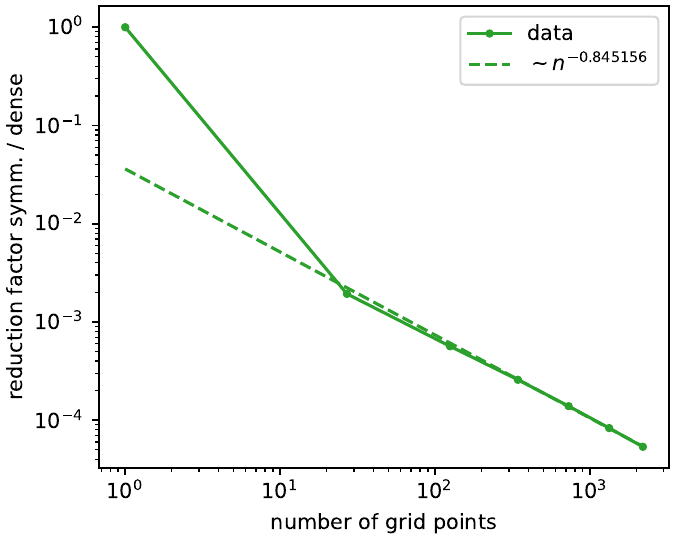}}
\caption{(a) Number of dense and symmetry-reduced ERI tensor entries as a function of the number of grid points $\abs{\Lambda_d}$, as listed in Table~\ref{tab:eri_num_entries}. (b) The reduction factor due to symmetries. The fitted curve indicates that the number of entries after symmetry reduction grows approximately as $\sim \abs{\Lambda_d}^{3.15}$.}
\label{fig:eri_num_entries}
\end{figure}

Table~\ref{tab:eri_num_entries} and Fig.~\ref{fig:eri_num_entries} list and visualize the number of dense and symmetry-reduced ERI tensor entries in dependence on the number of grid points $\abs{\Lambda_d}$, and the corresponding reduction factor. Solely based on translational invariance, one expects an asymptotic growth as $\sim \abs{\Lambda_d}^3$ (since there are on the order of $\abs{\Lambda_d}$ equivalent shifted versions of each ERI term). The observed slightly larger growth, despite the additional permutation and octahedral symmetries, is due to the fact that translations are restricted by the grid boundaries; for example, in the extreme case that the points $i, j, k, \ell$ are diagonally opposite corners of $\Lambda_d$, no translation is possible when confining all points to $\Lambda_d$.

Given a set of representative ERI indices, we can compute the actual integrals via a straightforward implementation of Eq.~\eqref{eq:coul_ijkl_computation}. A pseudocode version is shown in Algorithm~\ref{alg:compute_eri}, where we additionally truncate small coefficients based on a specified tolerance $\epsilon_{\text{tol}}$. We use the notation $x \addgets y$ as shorthand for $x \gets x + y$ (updating the variable $x$ by adding $y$ to it). Truncation by the threshold $\epsilon_{\text{tol}}$ is implemented by retaining only the factors $c_{\Delta i, \mu}$ for $\mu \in \{ M_{\Delta i}^{\min}, \dots, M_{\Delta i}^{\max} \}$ and $d_{\Delta i, \Delta k, \xi}$ for $\xi \in \{ N_{\Delta i, \Delta k}^{\min}, \dots, N_{\Delta i, \Delta k}^{\max} \}$, respectively.

\begin{algorithm}[H]
\caption{Compute ERIs}
\label{alg:compute_eri}
\begin{algorithmic}
\Require Gausslet coefficients $\{b_j\}_{j=-L,\dots,L}$, grid $\Lambda_d$, representative ERI indices $\text{repr\_list}$, truncation threshold $\epsilon_{\text{tol}}$
\State $\{c_{\Delta i}\} \gets$ \textproc{gausslet\_factors}($\{b_j\}$, $d$, $\epsilon_{\text{tol}}$)
\State $\{d_{\Delta i, \Delta k}\} \gets$ \textproc{eri\_factor\_products}($\{c_{\Delta i}\}$, $d$, $\epsilon_{\text{tol}}$)
\LineComment{fill look-up table for Coulomb integrals}
\For{$x = 0, 1, \dots, 3 \big(6 (d - 1) + 4 L\big)^2$}
    \State $\text{ctable}_x \gets C_{\frac{1}{3}}\big(\sqrt{x} / 6\big)$
\EndFor
\For{$(i, j, k, \ell) \in \text{repr\_list}$}
    \State $t_n \gets 3 (i_n + j_n - k_n - \ell_n)$ for $n = 1, 2, 3$
    \LineComment{compute coefficients in Eq.~\eqref{eq:eri_circular_contour_products_def}}
    \State initialize $f$ as a zero vector
    \For{$\xi_1 = N_{i_1 - j_1, k_1 - \ell_1}^{\min}, \dots, N_{i_1 - j_1, k_1 - \ell_1}^{\max}$}
        \If{$\abs{d_{i_1 - j_1, k_1 - \ell_1, \xi_1}} \le \epsilon_{\text{tol}}$}
            \State continue
        \EndIf
        \For{$\xi_2 = N_{i_2 - j_2, k_2 - \ell_2}^{\min}, \dots, N_{i_2 - j_2, k_2 - \ell_2}^{\max}$}
            \If{$\abs{d_{i_2 - j_2, k_2 - \ell_2, \xi_2}} \le \epsilon_{\text{tol}}$}
                \State continue
            \EndIf
            \State $f_{(t_1 + \xi_1)^2 + (t_2 + \xi_2)^2} \addgets$
            \State $\quad d_{i_1 - j_1, k_1 - \ell_1, \xi_1} d_{i_2 - j_2, k_2 - \ell_2, \xi_2}$
        \EndFor
    \EndFor
    \LineComment{evaluate Eq.~\eqref{eq:coul_ijkl_computation}}
    \State initialize $\text{val} \gets 0$
    \For{$\rho = 0, 1, \dots, \rho_{\max}$}
        \If{$f_{\rho} = 0$}
            \State continue
        \EndIf
        \For{$\xi_3 = N_{i_3 - j_3, k_3 - \ell_3}^{\min}, \dots, N_{i_3 - j_3, k_3 - \ell_3}^{\max}$}
            \If{$\abs{d_{i_3 - j_3, k_3 - \ell_3, \xi_3}} \le \epsilon_{\text{tol}}$}
                \State continue
            \EndIf
            \State $\text{val} \addgets f_{\rho} \, d_{i_3 - j_3, k_3 - \ell_3, \xi_3} \, \text{ctable}_{\rho + (t_3 + \xi_3)^2}$
        \EndFor
    \EndFor
    \State $\coul{ij \vert k\ell} \gets \text{val}$
\EndFor
\State \Return $\left\{ \coul{ij \vert k\ell} \,\vert\, (i, j, k, \ell) \in \text{repr\_list} \right\}$

\medskip

\Function{gausslet\_factors}{$\{b_j\}, d, \epsilon_{\text{tol}}$}
    \For{$\Delta i = -(d - 1), \dots, d - 1$}
        \LineComment{compute $c_{\Delta i, \mu}$ in Eq.~\eqref{eq:gausslet_factors}}
        \State initialize $c_{\Delta i, \mu} \gets 0$ for $\mu = -2 L, \dots, 2 L$
        \For{$\alpha = -L, \dots, L$}
            \For{$\beta = -L, \dots, L$}
                \State $c_{\Delta i, \alpha + \beta} \addgets \frac{\sqrt{\pi}}{3} b_{\alpha} b_{\beta} \e^{-\left(\frac{3 \Delta i + (\alpha - \beta)}{2}\right)^2}$
            \EndFor
        \EndFor
        \State $c_{\Delta i} \gets \{c_{\Delta i, \mu}\}_{\mu = M_{\Delta i}^{\min}, \dots, M_{\Delta i}^{\max}}$
        \LineComment{$M_{\Delta i}^{\min}$, $M_{\Delta i}^{\max}$ such that $\abs{c_{\Delta i, \mu}} \le \epsilon_{\text{tol}}$
        \LineComment for $\mu \in \{-2 L, \dots, 2 L\} \setminus \{ M_{\Delta i}^{\min}, \dots, M_{\Delta i}^{\max} \}$}
    \EndFor
    \State \Return $\{c_{\Delta i}\}_{\Delta i = -(d - 1), \dots, d - 1}$
\EndFunction

\medskip

\Function{eri\_factor\_products}{$\{c_{\Delta i}\}$, $d$, $\epsilon_{\text{tol}}$}
    \For{$\Delta i = -(d - 1), \dots, d - 1$}
        \For{$\Delta k = -(d - 1), \dots, d - 1$}
            \LineComment{compute $d_{\Delta i, \Delta k, \xi}$ in Eq.~\eqref{eq:eri_gausslet_factor_products_def}}
            \State initialize $d_{\Delta i, \Delta k, \xi} \gets 0$ for $\xi = -4 L, \dots, 4 L$
            \For{$\mu = M_{\Delta i}^{\min}, \dots, M_{\Delta i}^{\max}$}
                \For{$\nu = M_{\Delta k}^{\min}, \dots, M_{\Delta k}^{\max}$}
                \State $d_{\Delta i, \Delta k, \mu - \nu} \addgets c_{\Delta i, \mu} c_{\Delta k, \nu}$
                \EndFor
            \EndFor
            \State $d_{\Delta i, \Delta k} \gets \{d_{\Delta i, \Delta k, \xi}\}_{\xi = N_{\Delta i, \Delta k}^{\min}, \dots, N_{\Delta i, \Delta k}^{\max}}$
            \LineComment{$N_{\Delta i, \Delta k}^{\min}$, $N_{\Delta i, \Delta k}^{\max}$ such that $\abs{d_{\Delta i, \Delta k, \xi}} \le \epsilon_{\text{tol}}$
            \LineComment for $\xi \in \{-4 L, \dots, 4 L\} \setminus \{ N_{\Delta i, \Delta k}^{\min}, \dots, N_{\Delta i, \Delta k}^{\max} \}$}
        \EndFor
    \EndFor
    \State \Return $\{d_{\Delta i, \Delta k}\}_{\Delta i = -(d - 1), \dots, d - 1, \Delta k = -(d - 1), \dots, d - 1}$
\EndFunction
\end{algorithmic}
\end{algorithm}

\subsection{Diagonal electron repulsion integrals}

Our goal is to efficiently evaluate the overlap integrals in Eq.~\eqref{eq:v_ij_def} for the $\{ \mathcal{G}_i \}$ basis. As before, we use the formula \eqref{eq:coulomb_integral_formula} for $C_w(d)$. Substituting the definition \eqref{eq:gausslet_orbitals_def} together with the expansion \eqref{eq:gausslet_def} into Eq.~\eqref{eq:v_ij_def} and noting that $m_i = 1$ for all $i$ by construction of the Gausslets leads to
\begin{multline}
\mathbf{v}_{i,j} = \int_{\R^3} \int_{\R^3} \prod_{n=1}^3 \Bigg(\sum_{\alpha_n,\beta_n=-L}^L b_{\alpha_n} b_{\beta_n} \, g_{\alpha_n}(r_n - i_n)\\
\times g_{\beta_n}(r'_n - j_n) \Bigg) \frac{1}{\norm{r - r'}} \, \ud^3 r \, \ud^3 r'.
\end{multline}

Together with the definition \eqref{eq:coulomb_integral_def}, one obtains
\begin{multline}
\mathbf{v}_{i,j} = \left(\frac{\sqrt{2 \pi}}{3}\right)^6 \times\\
\sum_{\alpha_1,\beta_1=-L}^L b_{\alpha_1} b_{\beta_1}%
\sum_{\alpha_2,\beta_2=-L}^L b_{\alpha_2} b_{\beta_2}%
\sum_{\alpha_3,\beta_3=-L}^L b_{\alpha_3} b_{\beta_3}\\
\times C_{\sqrt{2}/3}\left(\norm*{\begin{pmatrix}i_n - j_n + \frac{\alpha_n - \beta_n}{3}\end{pmatrix}_{n=1,2,3}}\right).
\end{multline}
As before, we rearrange the summation to improve computational efficiency, noting that $C_{\sqrt{2}/3}$ only depends on the summation indices via $\alpha_n - \beta_n$. Let
\begin{equation}
\label{eq:erida_gausslet_factor_products_def}
c_{\mu} \coloneq \frac{2 \pi}{9} \sum_{\alpha,\beta=-L}^L b_{\alpha} b_{\beta} \, \delta_{\alpha - \beta, \mu}
\end{equation}
for $\mu \in \{-2 L, \dots, 2 L\}$. Then
\begin{multline}
\label{eq:v_ij_triple_sum}
\mathbf{v}_{i,j} = \sum_{\mu_1,\mu_2,\mu_3=-2L}^{2L} c_{\mu_1} \, c_{\mu_2} \, c_{\mu_3}\\
\times C_{\sqrt{2}/3}\left(\norm*{\begin{pmatrix}i_n - j_n + \frac{\mu_n}{3}\end{pmatrix}_{n=1,2,3}}\right).
\end{multline}
Similar to the ERI case, we can further reorganize the triple sum and precompute a look-up table for the Coulomb integral function $C_{\sqrt{2}/3}$ by mapping the argument of $C_{\sqrt{2}/3}$ to an integer: we multiply the inner vector by $3$ and square its norm:
\begin{multline}
C_{\sqrt{2}/3}\left(\norm*{\begin{pmatrix}i_n - j_n + \frac{\mu_n}{3}\end{pmatrix}_{n=1,2,3}}\right) \\
= \tilde{C}_{\sqrt{2}/3}\left(\sum_{n=1}^3 \big(3 (i_n - j_n) + \mu_n\big)^2\right)
\end{multline}
with
\begin{equation}
\tilde{C}_{\sqrt{2}/3}(x) \coloneq C_{\sqrt{2}/3}(\sqrt{x} / 3), \quad x \in \R_{\ge 0}.
\end{equation}
This allows us to reorganize the summations in Eq.~\eqref{eq:v_ij_triple_sum}: introduce (for $t_1, t_2 \in \Z$ and $\rho \in \N_0$)
\begin{equation}
\label{eq:erida_circular_contour_products_def}
f^{t_1, t_2}_{\rho} \coloneq \sum_{\mu_1,\mu_2=-2L}^{2L} c_{\mu_1} c_{\mu_2} \, \delta_{(t_1 + \mu_1)^2 + (t_2 + \mu_2)^2, \rho},
\end{equation}
then
\begin{equation}
\label{eq:v_ij_computation}
\mathbf{v}_{i,j} =%
\sum_{\rho=0}^{\rho_{\max}} f^{t_1, t_2}_{\rho} \sum_{\mu_3=-2L}^{2L} c_{\mu_3} \, \tilde{C}_{\sqrt{2}/3}\left(\rho + (t_3 + \mu_3)^2\right)
\end{equation}
with $\rho_{\max} = (\abs{t_1} + 2 L)^2 + (\abs{t_2} + 2 L)^2$ and $t_n = 3 (i_n - j_n)$ for $n = 1, 2, 3$.

As above, we can further reduce the computational cost by filtering out coefficients $\abs{c_{\mu}} \le \epsilon_{\text{tol}}$.

Analogous to the ERIs, we exploit symmetries by first listing a minimal representative set of grid index tuples $(i, j) \in \Lambda_d^2$ such that all remaining ERIDA integrals (on $\Lambda_d$) can be obtained from this set by symmetry transformations of the grid indices. The equivalence classes are determined via the equivalence relation (for $i, j, i', j' \in \Lambda_d$):
\begin{equation}
(i, j) \sim (i', j')
\end{equation}
precisely if $(i, j)$ can be mapped to $(i', j')$ by a translation, index flip $i \leftrightarrow j$, and/or octahedral point group transformation as described in Sec.~\ref{sec:symmetries}.

\begin{table}
\begin{tabular}{|crrrl|}
\hline
$\Lambda_d$ dim. & $\abs{\Lambda_d}$ &  \# dense & \# symm. & reduction \\
                 &                   &     ERIDA &    ERIDA &           \\
\hline
$1 \times 1 \times 1$    &         1 &         1 &        1 & 1         \\
$3 \times 3 \times 3$    &        27 &       729 &       10 & 0.014     \\
$5 \times 5 \times 5$    &       125 &     15625 &       35 & 0.0022    \\
$7 \times 7 \times 7$    &       343 &    117649 &       84 & 0.00071   \\
$9 \times 9 \times 9$    &       729 &    531441 &      165 & 0.00031   \\
$11 \times 11 \times 11$ &      1331 &   1771561 &      286 & 0.00016   \\
$13 \times 13 \times 13$ &      2197 &   4826809 &      455 & 0.000094  \\
$15 \times 15 \times 15$ &      3375 &  11390625 &      680 & 0.000060  \\
$17 \times 17 \times 17$ &      4913 &  24137569 &      969 & 0.000040  \\
$19 \times 19 \times 19$ &      6859 &  47045881 &     1330 & 0.000028  \\
$21 \times 21 \times 21$ &      9261 &  85766121 &     1771 & 0.000020  \\
$23 \times 23 \times 23$ &     12167 & 148035889 &     2300 & 0.000016  \\
$25 \times 25 \times 25$ &     15625 & 244140625 &     2925 & 0.000012  \\
$27 \times 27 \times 27$ &     19683 & 387420489 &     3654 & 0.0000094 \\
\hline
\end{tabular}
\caption{Number of dense and symmetry-reduced ERIDA tensor entries for increasing grid dimensions, analogous to Table~\ref{tab:eri_num_entries}. The number of dense entries in the third column equals $\abs{\Lambda_d}^2 = d^6$. The fourth column shows the number of integral evaluations actually required due to symmetry reduction, adhering to the formula $\binom{d + 2}{3}$, and the last column contains the corresponding reduction factor, scaling asymptotically as $\sim 1 / (6 d^3)$.}
\label{tab:erida_num_entries}
\end{table}

For the sake of completeness, the enumeration of representative ERIDA indices is described in algorithm~\ref{alg:enumerate_representative_erida} and the evaluation of these integrals in algorithm~\ref{alg:compute_erida}.

\begin{algorithm}[H]
\caption{Enumerate representative ERIDA indices}
\label{alg:enumerate_representative_erida}
\begin{algorithmic}
\Require $\Lambda_d$ for odd positive integer $d$
\State $s_d \gets \frac{1}{2} (d - 1)$
\State $\text{repr\_list} \gets \emptyset$
\For{$i_1, j_1 \in \left\{ -s_d, \dots, s_d \right\}$}
    \LineComment{$x$-coordinate of bounding box center}
    \State $b_1 \gets \frac{1}{2}(i_1 + j_1)$
    \If{$b_1 \notin \{-\frac{1}{2}, 0, \frac{1}{2}\}$}
        \State continue
    \EndIf
    \For{$i_2, j_2 \in \left\{ -s_d, \dots, s_d \right\}$}
        \LineComment{$y$-coordinate of bounding box center}
        \State $b_2 \gets \frac{1}{2}(i_2 + j_2)$
        \If{$b_2 \notin \{-\frac{1}{2}, 0, \frac{1}{2}\}$}
            \State continue
        \EndIf
        \For{$i_3, j_3 \in \left\{ -s_d, \dots, s_d \right\}$}
            \LineComment{$z$-coordinate of bounding box center}
            \State $b_3 \gets \frac{1}{2}(i_3 + j_3)$
            \If{$b_3 \notin \{-\frac{1}{2}, 0, \frac{1}{2}\}$}
                \State continue
            \EndIf
            \State $\text{idx} \gets \gridindex{d}(i) \, \abs{\Lambda_d} + \gridindex{d}(j)$
            \State is\_minimal\_index $\gets$ true
            \For{$\sigma_1, \sigma_2, \sigma_3 \in \{ 0, 1 \}$}
                \LineComment{translation bias}
                \State $\text{bias} \gets (2 \sigma_1 b_1, 2 \sigma_2 b_2, 2 \sigma_3 b_3)$
                 \If{not \begin{varwidth}[t]{0.5\linewidth}\textproc{is\_min\_orbit\_erida\_index}($\Lambda_d$, $(i - \text{bias}, j - \text{bias})$, idx)\end{varwidth}}
                    \State is\_minimal\_index $\gets$ false
                \EndIf
            \EndFor
            \If{is\_minimal\_index}
                \State add $(i, j)$ to $\text{repr\_list}$
            \EndIf
        \EndFor
    \EndFor
\EndFor
\State \Return repr\_list

\medskip

\Function{is\_min\_orbit\_erida\_index}{$\Lambda_d$, $(i, j)$, $\text{idx}$}
    \LineComment{iterate over the 48 group elements of $\Oh$}
    \For{$u \in \Oh$}
    \State $(i', j') \gets \left(u(i), u(j)\right)$
        \If{$\gridindex{d}(i') > \gridindex{d}(j')$}
            \State swap $i' \leftrightarrow j'$
        \EndIf
        \If{$\gridindex{d}(i') \, \abs{\Lambda_d} + \gridindex{d}(j') < \text{idx}$}
            \State \Return false
        \EndIf
    \EndFor
    \State \Return true
\EndFunction
\end{algorithmic}
\end{algorithm}

Table~\ref{tab:erida_num_entries} lists the number of dense and symmetry-reduced ERIDA tensor entries as a function of the number of grid points $\abs{\Lambda_d}$, and the corresponding reduction factor. One notices that the number of integral evaluations after symmetry reduction is even smaller than the number of grid points (except for the trivial $1 \times 1 \times 1$ grid supporting a single integral). Indeed, the number of equivalence classes adheres to the closed form expression $\binom{d + 2}{3}$: two tuples $(i, j)$ and $(i', j')$ are equivalent precisely when the difference vectors $i-j$ and $i'-j'$ lie on the same $\Oh$ orbit up to sign, i.e., the multisets $\{\abs{i_1 - j_1}, \abs{i_2 - j_2}, \abs{i_3 - j_3}\}$ coincide, and there are $\binom{d + 2}{3}$ such multisets with values taken from $\{0, \dots, d-1\}$. Consequently, the reduction factor equals $\binom{d + 2}{3} / d^6 \sim 1 / (6 d^3)$ for large $d$. In other words, around six times fewer integral evaluations than the number of grid points are required as $d \to \infty$.

\begin{algorithm}[H]
\caption{Compute ERIDAs}
\label{alg:compute_erida}
\begin{algorithmic}
\Require Gausslet coefficients $\{b_j\}_{j=-L,\dots,L}$, grid $\Lambda_d$, representative ERIDA indices $\text{repr\_list}$, truncation threshold $\epsilon_{\text{tol}}$
\State $\{c_{\mu}\} \gets$ \textproc{erida\_factor\_products}($\{b_j\}$, $d$, $\epsilon_{\text{tol}}$)
\LineComment{fill look-up table for Coulomb integrals}
\For{$x = 0, 1, \dots, 3 \big(3 (d - 1) + 2 L\big)^2$}
    \State $\text{ctable}_x \gets C_{\sqrt{2}/3}\big(\sqrt{x} / 3\big)$
\EndFor
\For{$(i, j) \in \text{repr\_list}$}
    \State $t_n = 3 (i_n - j_n)$ for $n = 1, 2, 3$
    \LineComment{compute coefficients in Eq.~\eqref{eq:erida_circular_contour_products_def}}
    \State initialize $f$ as a zero vector
    \For{$\mu_1 = M^{\min}, \dots, M^{\max}$}
        \If{$\abs{c_{\mu_1}} \le \epsilon_{\text{tol}}$}
            \State continue
        \EndIf
        \For{$\mu_2 = M^{\min}, \dots, M^{\max}$}
            \If{$\abs{c_{\mu_2}} \le \epsilon_{\text{tol}}$}
                \State continue
            \EndIf
            \State $f_{(t_1 + \mu_1)^2 + (t_2 + \mu_2)^2} \addgets c_{\mu_1} c_{\mu_2}$
        \EndFor
    \EndFor
    \LineComment{evaluate Eq.~\eqref{eq:v_ij_computation}}
    \State initialize $\text{val} \gets 0$
    \For{$\rho = 0, 1, \dots, \rho_{\max}$}
        \If{$f_{\rho} = 0$}
            \State continue
        \EndIf
        \For{$\mu_3 = M^{\min}, \dots, M^{\max}$}
            \If{$\abs{c_{\mu_3}} \le \epsilon_{\text{tol}}$}
                \State continue
            \EndIf
            \State $\text{val} \addgets f_{\rho} \, c_{\mu_3} \, \text{ctable}_{\rho + (t_3 + \mu_3)^2}$
        \EndFor
    \EndFor
    \State $\mathbf{v}_{i,j} \gets \text{val}$
\EndFor
\State \Return $\mathbf{v}$

\medskip

\Function{erida\_factor\_products}{$\{b_j\}, d, \epsilon_{\text{tol}}$}
    \LineComment{compute $c_{\mu}$ in Eq.~\eqref{eq:erida_gausslet_factor_products_def}}
    \State initialize $c_{\mu} \gets 0$ for $\mu = -2 L, \dots, 2 L$
    \For{$\alpha = -L, \dots, L$}
        \For{$\beta = -L, \dots, L$}
            \State $c_{\alpha - \beta} \addgets \frac{2 \pi}{9} b_{\alpha} b_{\beta}$
        \EndFor
    \EndFor
    \State determine $M^{\min}$, $M^{\max}$ such that $\abs{c_{\mu}} \le \epsilon_{\text{tol}}$
    \State for $\mu \in \{-2 L, \dots, 2 L\} \setminus \{ M^{\min}, \dots, M^{\max} \}$
    \State \Return $\{c_{\mu}\}_{\mu = M^{\min}, \dots, M^{\max}}$
\EndFunction
\end{algorithmic}
\end{algorithm}

\section{Numerical experiments}

As a demonstration and test of our implementation, we use the developed functionality of evaluating overlap integrals to study their asymptotic behavior and to compute the (approximate) ground states of the hydrogen atom and molecule.

\subsection{Asymptotic decay of the diagonal electron repulsion integrals}

\begin{figure}[!ht]
\centering
\subfloat[integral values]{\includegraphics[width=0.9\columnwidth]{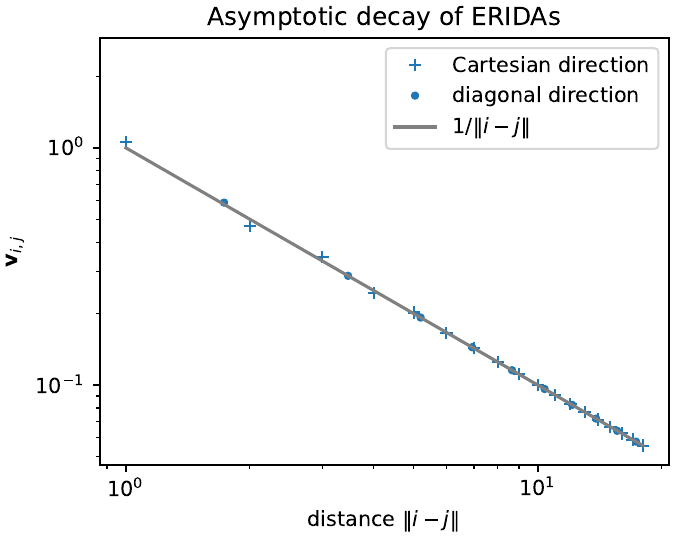}}\\
\subfloat[deviation from 1/distance]{\includegraphics[width=0.9\columnwidth]{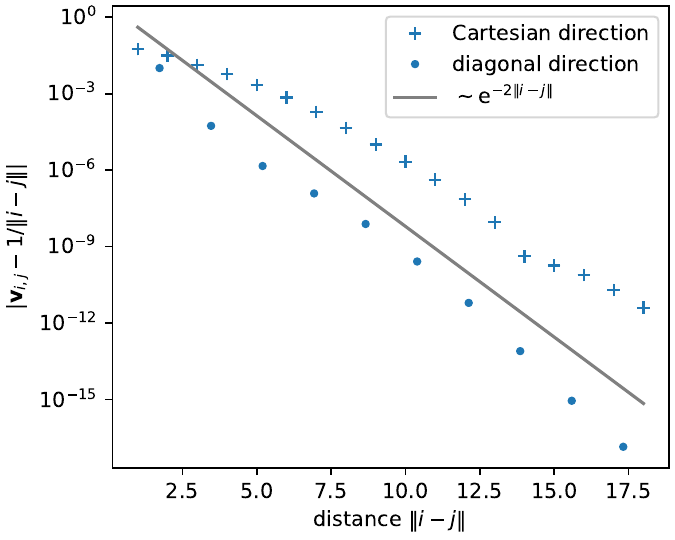}}
\caption{Asymptotic decay behavior of the electron repulsion integrals based on the integral diagonal approximation. The Cartesian and diagonal directions refer to difference vectors of the form $i - j = (\ell, 0, 0)$ and $i - j = (\ell, \ell, \ell)$, respectively, for $\ell \in \Z$. (a) shows the integral values on a log-log scale, and (b) the deviation from $1 / \norm{i - j}$ on a semilogarithmic scale.}
\label{fig:erida_integrals_asymptotic_decay}
\end{figure}

The localization and exponential decay of the Gausslet orbitals imply a delta-function-type behavior at large distances. Accordingly, the ERIDA overlap integrals are expected to scale as the inverse distance between the respective grid points. We quantify this observation in Fig.~\ref{fig:erida_integrals_asymptotic_decay}, showing the integral values for increasing grid point distances in the top subfigure, and the deviation from the inverse distance, i.e.,
\begin{equation}
\mathrm{deviation}(i - j) \coloneq \abs*{\mathbf{v}_{i,j} - \frac{1}{\norm{i - j}}} \quad \text{for } i, j \in \Lambda
\end{equation}
in the bottom subfigure. The Cartesian and diagonal directions refer to difference vectors of the form $i - j = (\ell, 0, 0)$ and $i - j = (\ell, \ell, \ell)$, respectively, for $\ell \in \Z$. As expected based on the exponential decay of the Gausslet orbitals, one observes an analogous exponential decay of this deviation. We heuristically find a  $\sim \mathrm{e}^{-2 \norm{i - j}}$ scaling.

\subsection{Hydrogen atom}

\begin{figure}[!ht]
\centering
\subfloat[ground state energy]{\includegraphics[width=\columnwidth]{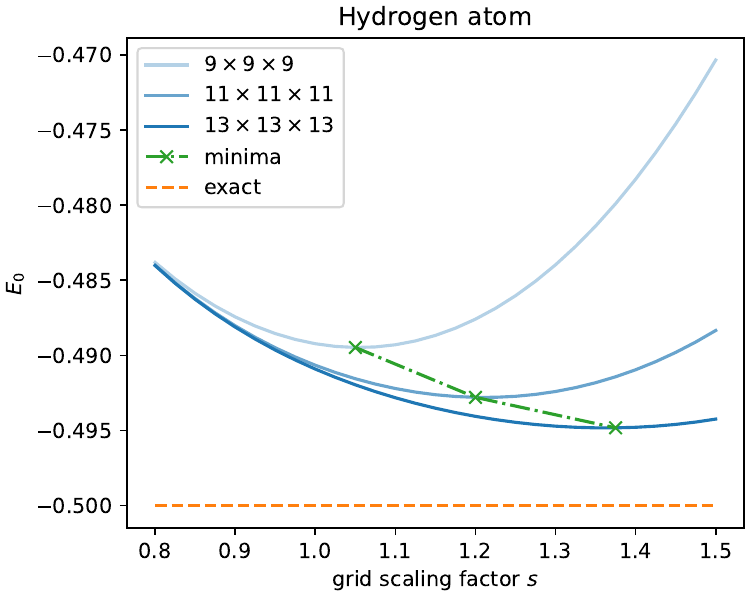}\label{fig:hydrogen_atom_energy}}\\
\subfloat[ground state wavefunction]{\includegraphics[width=\columnwidth]{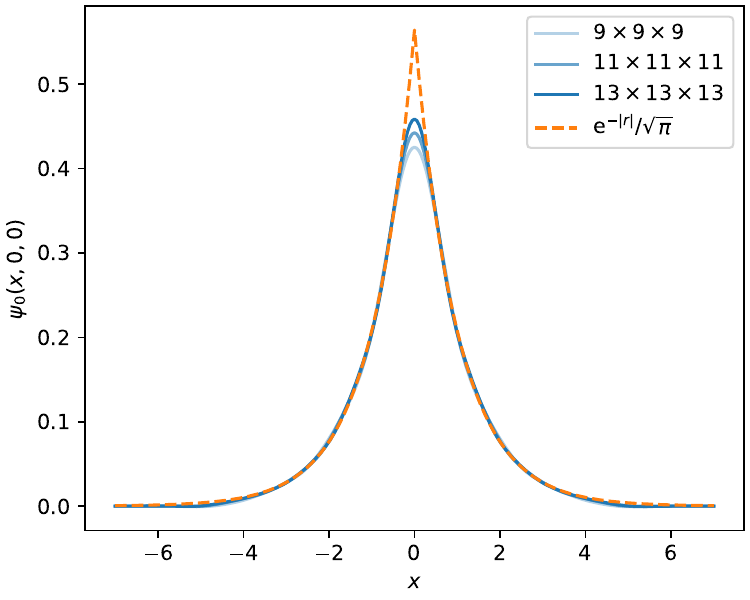}}
\caption{Numerical approximation of the ground state energy and the corresponding wavefunction using Gausslet orbitals, for increasing grid length $d$ and as a function of the grid scaling factor $s$, see Sec.~\ref{sec:rescaling}. The calculations use atomic units. The orange dashed curve shows the analytically exact solution. The green markers in (a) denote the respective minima of each curve; the wavefunctions shown in (b) are the corresponding eigenstates at the minimal energies and are evaluated in real space along the path $(x, 0, 0)$ for $x \in \R$.}
\label{fig:hydrogen_atom}
\end{figure}

We consider the hydrogen atom as a single-electron system with an exact analytic solution, serving as a simple benchmark demonstration of the Gausslet orbitals. The Hamiltonian in atomic units reads
\begin{equation}
H_{\text{ha}} = -\frac{1}{2} \Delta - \frac{1}{\norm{r}}
\end{equation}
where $\Delta$ denotes the Laplace operator, acting on square-integrable single-electron wavefunctions $\psi: \R^3 \to \C$. Note the absence of electron-electron repulsion terms. The Gausslet orbital basis on a finite grid corresponds to the Ansatz
\begin{equation}
\psi(r) = \sum_{i \in \Lambda_d} \psi_i \, s^{3/2} \mathcal{G}_i(s r), \quad r \in \R^3
\end{equation}
with $s > 0$ the orbital (or grid) rescaling factor (discussed in Sec.~\ref{sec:rescaling}) and $\{\psi_i\}_{i \in \Lambda_d}$ the to-be-determined coefficient vector, which can be chosen real-valued for the ground state problem. We use exact diagonalization, specifically NumPy's \texttt{eigh} function, of
\begin{equation}
\mathbf{h}_{\text{ha}} \coloneq \left(s^2 \, \mathbf{k}_{i, j} - s \, \mathfrak{n}_{i, j}^1(0)\right)_{i,j \in \Lambda_d}
\end{equation}
to compute the ground-state energy and corresponding coefficient vector.

Fig.~\ref{fig:hydrogen_atom} shows the obtained ground state energies and corresponding wavefunctions for increasing grid sizes $\Lambda_d$ and as a function of $s$. Recall that increasing $s$ means a finer grid. As the number of available grid points in each coordinate direction increases, the optimal energy is obtained by refining the grid (while still capturing the bulk of the wavefunction) --- in agreement with the green markers in Fig.~\ref{fig:hydrogen_atom_energy} moving towards larger $s$. The minimal energy value for the $13 \times 13 \times 13$ grid reads $-0.4948287\,\mathrm{hartree}$. One observes a good agreement between the Gausslet-based wavefunction and the analytic reference, except for the Gausslets' inability to fully resolve the cusp at the origin, which is attributable to their smooth outline for the given grid resolution.

\subsection{Hydrogen molecule}

A minimal system where the electron-electron repulsion becomes relevant is the hydrogen molecule, governed by the Hamiltonian (in atomic units)
\begin{multline}
H_{\text{hm}} = -\frac{1}{2} \Delta_{r} -\frac{1}{2} \Delta_{r'} \\
- \frac{1}{\norm{r - R}} - \frac{1}{\norm{r - R'}} - \frac{1}{\norm{r' - R}} - \frac{1}{\norm{r' - R'}} \\
+ \frac{1}{\norm{r - r'}} + \frac{1}{\norm{R - R'}}
\end{multline}
where $R$ and $R'$ denote the respective positions of the nuclei. For the following calculations, they are located at $(\pm \frac{1}{2} \ell_{\text{hm}}, 0, 0)$ with $\ell_{\text{hm}} = 1.4\,\mathrm{bohr}$ denoting the (equilibrium) nuclear distance. The nuclear-nuclear repulsion (last term of the Hamiltonian) acts as a multiplicative constant on the electronic wavefunction and as a shift of the total energy.

Anticipating that the hydrogen molecule ground state is a spin singlet, we use the wavefunction Ansatz
\begin{equation}
\ket{\psi} = \psi(r, r') \, \frac{1}{\sqrt{2}}(\ket{\uparrow\downarrow} - \ket{\downarrow\uparrow}),
\end{equation}
where $r, r' \in \R^3$ are the spatial coordinates of the two electrons, and the spatial wavefunction is \emph{symmetric}: $\psi(r, r') = \psi(r', r)$. Analogous to the hydrogen atom, the Gausslet orbital basis on a finite grid corresponds to the representation
\begin{equation}
\label{eq:two_electron_wavefunction}
\psi(r, r') = \sum_{i, i' \in \Lambda_d} \psi_{i,i'} \, s^3 \mathcal{G}_i(s r) \mathcal{G}_{i'}(s r'), \quad r, r' \in \R^3
\end{equation}
with $s > 0$ the orbital (or grid) rescaling factor. The task consists of determining the (real-valued) coefficient vector $\{\psi_{i,i'}\}_{i, i' \in \Lambda_d}$ under the symmetry constraint $\psi_{i,i'} = \psi_{i',i}$ for all $i, i' \in \Lambda_d$.

On the Hamiltonian level, the Gausslet orbital discretization leads to the one-body matrix
\begin{multline}
\mathbf{h}_{\text{hm}}^{(1)} \coloneq \Big(s^2 \, \mathbf{k}_{i, j} - s \, \mathfrak{n}_{i,j}^1\big((- s \tfrac{1}{2} \ell_{\text{hm}}, 0, 0)\big) \\
- s \, \mathfrak{n}_{i,j}^1\big((s \tfrac{1}{2} \ell_{\text{hm}}, 0, 0)\big)\Big)_{i,j \in \Lambda_d},
\end{multline}
the two-body matrix (note the index ordering)
\begin{equation}
\mathbf{h}_{\text{hm,ERI}}^{(2)} \coloneq \Big(\coul{ij \vert k\ell}\Big)_{(i,k), (j,\ell) \in \Lambda_d^2}
\end{equation}
and the overall Hamiltonian
\begin{equation}
\mathbf{h}_{\text{hm,ERI}} = \mathbf{h}_{\text{hm}}^{(1)} \otimes \mathds{1} + \mathds{1} \otimes \mathbf{h}_{\text{hm}}^{(1)} + \mathbf{h}_{\text{hm,ERI}}^{(2)} \in \R^{d^6 \times d^6}.
\end{equation}
The energy contribution of the nuclear-nuclear repulsion $1 / \ell_{\text{hm}}$ is added as a post-processing step. For computational efficiency, we do not explicitly evaluate the above Kronecker products, but instead implement the action of the one-body matrices on a statevector as
\begin{equation}
\label{eq:h_hm_1_application}
\big(\mathbf{h}_{\text{hm}}^{(1)} \otimes \mathds{1} + \mathds{1} \otimes \mathbf{h}_{\text{hm}}^{(1)}\big) \psi = \mathbf{h}_{\text{hm}}^{(1)} \cdot \psi_{\text{mat}} + \psi_{\text{mat}} \cdot \big(\mathbf{h}_{\text{hm}}^{(1)}\big)^T,
\end{equation}
i.e., as matrix-matrix multiplications when temporarily interpreting $\psi$ as a matrix. The steep memory cost of storing the dense ERI tensor (after $j \leftrightarrow k$ index permutation) in $\mathbf{h}_{\text{hm,ERI}}^{(2)}$ limits the following demonstrations to relatively small grid dimensions. Under the integral diagonal approximation, the two-body term becomes
\begin{equation}
\mathbf{h}_{\text{hm,ERIDA}}^{(2)} \coloneq \Big(\delta_{ij} \delta_{k\ell} \, \mathbf{v}_{i,k}\Big)_{(i,k), (j,\ell) \in \Lambda_d^2}.
\end{equation}
Its action on the quantum state is an (efficiently computable) pointwise multiplication:
\begin{equation}
\label{eq:h_hm2_erida_pointwise}
\mathbf{h}_{\text{hm,ERIDA}}^{(2)} \psi = \mathbf{v} \odot \psi_{\text{mat}},
\end{equation}
where $\odot$ denotes the Hadamard product. This form reflects the original real-space formulation $\frac{1}{\norm{r - r'}} \psi(r, r')$, which is likewise a pointwise multiplication. We use a sparse eigenvalue solver, specifically SciPy's \texttt{sparse.linalg.eigsh}, to compute the ground-state energies and corresponding eigenvectors of $\mathbf{h}_{\text{hm,ERI}}$ and $\mathbf{h}_{\text{hm,ERIDA}}$, exploiting the ``matrix-free'' application of the latter in Eqs.~\eqref{eq:h_hm_1_application} and \eqref{eq:h_hm2_erida_pointwise}.

\begin{figure}[!ht]
\centering
\subfloat[full ERI, small grid dimensions]{\includegraphics[width=\columnwidth]{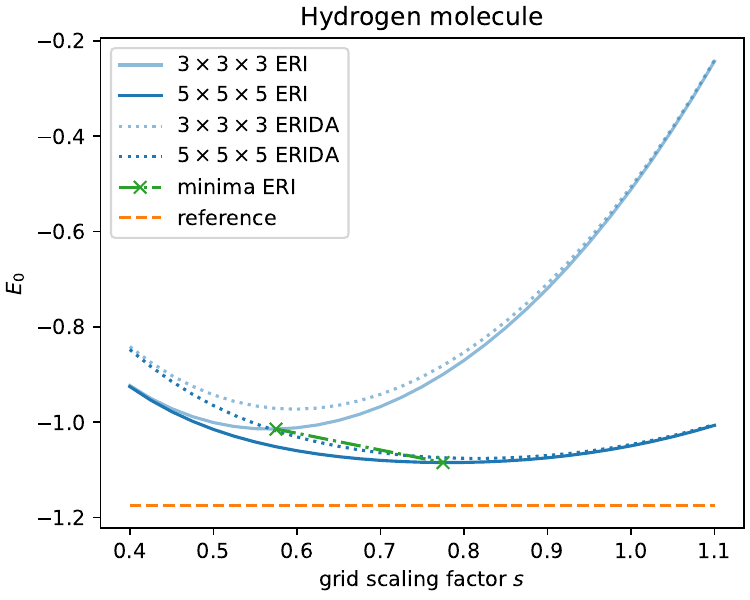}\label{fig:hydrogen_molecule_eri_energy}}\\
\subfloat[ERIDA, larger grid dimensions]{\includegraphics[width=\columnwidth]{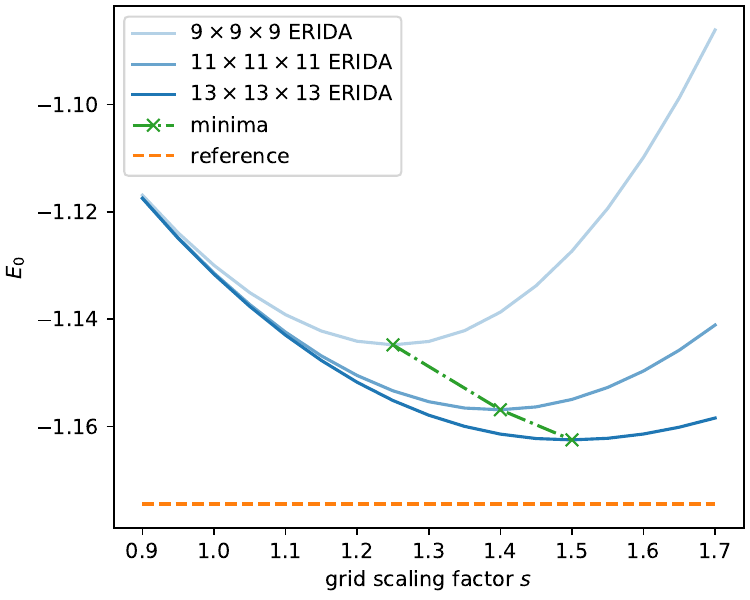}\label{fig:hydrogen_molecule_erida_energy}}
\caption{Numerical approximation of the ground state energy of the hydrogen molecule at equilibrium nuclear distance $1.4\,\mathrm{bohr}$ using Gausslet orbitals, as a function of the grid scaling factor $s$. The orange dashed line shows the reference value from \cite{Sims2006}. In (a), the solid curves are based on the full ERI tensor, and the dotted curves show calculations based on the integral diagonal approximation for comparison. (b) shows the energy curves obtained with the integral diagonal approximation for larger grid dimensions. Note that the two subfigures use different axes ranges.}
\label{fig:hydrogen_molecule_energy}
\end{figure}

The obtained energy curves are plotted in Fig~\ref{fig:hydrogen_molecule_energy} as a function of the grid scaling factor $s$. The reference value $-1.174475714220\,\mathrm{hartree}$ (dashed orange line) is taken from \cite{Sims2006}. The subfigure~\ref{fig:hydrogen_molecule_eri_energy} compares the results with and without the integral diagonal approximation (and otherwise identical parameters). Note that the variational principle (the numerically computed energy is lower-bounded by the true ground-state energy) holds, in a strict sense, only for the ERI calculation. One observes that the ERIDA energy curves lie above the ERI curves, and that the deviation decreases as $s$ increases. Subfigure~\ref{fig:hydrogen_molecule_erida_energy} visualizes the energy curves for larger grid dimensions, which become accessible under the integral diagonal approximation. The smallest computed energy value is $-1.162519\,\mathrm{hartree}$, which is thus about $12\,\mathrm{millihartrees}$ above the reference value.

\begin{figure}[!ht]
\centering
\includegraphics[width=\columnwidth]{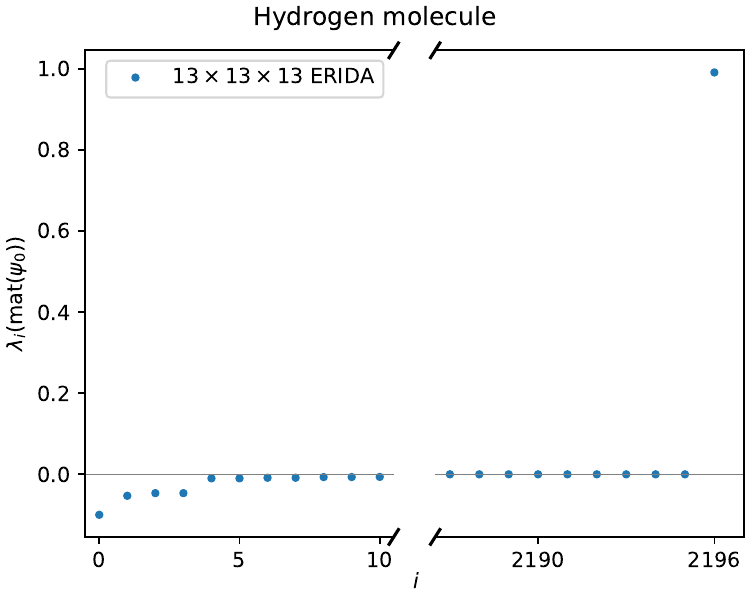}
\caption{Eigenvalues of the hydrogen molecule ground state coefficient vector $\psi_{i,i'}$ (see Eq.~\eqref{eq:two_electron_wavefunction}) interpreted as a matrix, using Gausslets on a $13 \times 13 \times 13$ grid and the integral diagonal approximation. The eigenvalues are listed in ascending order. Note that the figure only shows the leading and trailing eigenvalues; the remaining ones in the ``bulk'' of the spectrum are close to zero.}
\label{fig:hydrogen_molecule_erida_wavefunction_spectrum}
\end{figure}

Insights about the ground state wavefunction can be obtained by interpreting the coefficient vector $\psi_{i,i'}$ as a symmetric matrix and computing its eigenvalues. Fig.~\ref{fig:hydrogen_molecule_erida_wavefunction_spectrum} visualizes these for the ground state on the $13 \times 13 \times 13$ grid. The numerical values are $(-0.0998, -0.0529, -0.0464, \dots, -3.17 \cdot 10^{-12}, 0.991)$. Their absolute values correspond to the singular values of the Schmidt decomposition. One notices a few negative eigenvalues with small magnitude, a sequence of almost-zero eigenvalues, and a dominating eigenvalue close to $1$. Thus, $\psi_{i,i'}$ is a low-rank matrix that could be well approximated by a few product states.

\section{Conclusions and outlook}

The location of the Gausslets on cubic lattice points admits an extensive exploitation of symmetries and a precomputation of terms along circular contours for the ERIs and ERIDAs, see Eqs.~\eqref{eq:eri_circular_contour_products_def} and \eqref{eq:erida_circular_contour_products_def}. These approaches for reducing computational cost are only partially applicable to nuclear overlap integrals, since the nuclei need not be positioned on lattice points. Indeed, in our numerical experiments, the nuclear integral computation requires the most wall-clock time among the overlap integrals. An approach to mitigate this issue could be interpolation, i.e., precomputing overlap integrals for nuclei at lattice points and interpolating integral values for general nuclear positions. We leave an exploration of this idea for future work.

Reaching large grid dimensions, on the order of hundreds or thousands of points in each coordinate direction, requires additional algorithmic innovations, since the computational cost and memory requirements of obtaining all overlap integrals on such grids are prohibitively large. A natural idea is the employment of fast multipole methods \cite{GreengardRokhlin1987} to avoid the $\mathcal{O}(\abs{\Lambda_d}^2)$ scaling when ``naively'' enumerating all pairwise interactions. Multipole methods align well with the localization of the Gausslets and corresponding $\sim \frac{1}{\norm{r}}$ (Coulomb) scaling of the nuclear and electron repulsion integrals in the far field, as we have quantified in Fig.~\ref{fig:erida_integrals_asymptotic_decay}.

Gausslets on a cubic lattice offer the advantage of a smooth wavefunction Ansatz at the trade-off of a spatial resolution limit given by the grid spacing. A promising approach to overcome this limitation in electronic structure simulations could be the inclusion of a Jastrow factor in the wavefunction Ansatz (see the ``transcorrelated electrons method'' \cite{Alavi2019, Lee2023}), or pseudopotential approaches, which combine the nucleus and core electrons into a smoothed effective potential. We also like to mention recent follow-up work on Gausslets \cite{WhiteStoudenmire2019, WhiteLindsey2023Nested, White2026Radial, White2026Angular} for non-uniform grids, in particular radial Gausslets. Developing efficient algorithms to compute overlap integrals for such variants suggests a follow-up project of the present work.

From a broader perspective, the Gausslet orbitals can be regarded as a ``first quantization'' formulation, in the sense of interpreting the lattice points as a discretization of the electron positions. Simulating larger systems (from tens to hundreds of electrons) poses the challenge of constructing a wavefunction representation capable of handling grids with thousands of points. This could be realized by assembling a smaller ``molecular orbital'' basis set from the Gausslets, or via dedicated tensor network Ans\"atze, such as projected entangled pair states (PEPS) in three dimensions or locality-preserving (shallow) quantum circuit states. We explore such quantum circuit states in an ongoing work.

\acknowledgments

We thank Garnet Chan for helpful discussions. This research is part of the Munich Quantum Valley, which is supported by the Bavarian state
government with funds from the Hightech Agenda Bayern Plus.

\bibliography{references}

\end{document}

%% file: figures/translation.tikz
\begin{tikzpicture}[>=stealth, scale=0.75]
% grid points
\foreach \y in {-5, ..., 5}
{
    \foreach \x in {-5, ..., 5}
    {
        \draw[fill] (\x, \y) circle (0.05);
    }
}
% bounding boxes
\draw ( 1, 1) rectangle ( 5, 4);
\draw (-2,-1) rectangle ( 2, 2);
% original points
\draw ( 4, 2) circle (0.1);  % i
\draw ( 5, 4) circle (0.1);  % j
\draw ( 3, 1) circle (0.1);  % k
\draw ( 1, 3) circle (0.1);  % l
% translated points
\draw ( 1, 0) circle (0.1);  % i'
\draw ( 2, 2) circle (0.1);  % j'
\draw ( 0,-1) circle (0.1);  % k'
\draw (-2, 1) circle (0.1);  % l'
\node at ( 4,   2.3) {\footnotesize $i$};
\node at ( 5.3, 4  ) {\footnotesize $j$};
\node at ( 3,   0.7) {\footnotesize $k$};
\node at ( 0.7, 3  ) {\footnotesize $\ell$};
\node at ( 1,   0.3) {\footnotesize $i'$};
\node at ( 2.3, 2  ) {\footnotesize $j'$};
\node at ( 0,  -1.3) {\footnotesize $k'$};
\node at (-2.3, 1  ) {\footnotesize $\ell'$};
\draw[->, dashed, blue] ( 1, 1) -- (-2,-1);
\draw[->, dashed, blue] ( 5, 1) -- ( 2,-1);
\draw[->, dashed, blue] ( 5, 4) -- ( 2, 2);
\draw[->, dashed, blue] ( 1, 4) -- (-2, 2);
% origin
\node at ( 0, -0.25) {\footnotesize $0$};
\end{tikzpicture}

%% file: figures/eri_dihedral_symmetry.tikz
\begin{tikzpicture}[>=stealth, scale=0.75]
\draw[dotted] (-1.5, 0) -- ( 1.5, 0);
\draw[dotted] (0, -1.5) -- ( 0, 1.5);
\draw[dotted] (-1.4, -1.4) -- ( 1.4, 1.4);
\draw[dotted] (-1.4,  1.4) -- ( 1.4,-1.4);
\draw[->] (0.4*0.8660254037844387, 0.4*0.5) arc [start angle=30, end angle=120, radius=0.4];
\draw (-1,-1) rectangle ( 1, 1);
\node[circle, fill=white, inner sep=1.5] at (-1, 1) {\footnotesize $i$};
\node[circle, fill=white, inner sep=1.5] at ( 1, 1) {\footnotesize $k$};
\node[circle, fill=white, inner sep=1.5] at (-1,-1) {\footnotesize $\ell$};
\node[circle, fill=white, inner sep=1.5] at ( 1,-1) {\footnotesize $j$};
\end{tikzpicture}